\documentclass[twocolumn]{aastex61}
\usepackage{amssymb}
\usepackage{amsmath}
\usepackage{relsize}		
\usepackage{framed}        

\renewcommand{\leq}{\leqslant}

\newcommand{\Msun}{\mathrm{M}_\odot}         
\newcommand{\Zsun}{\mathrm{Z}_\odot}         
\newcommand{\HII}{H{\relsize{-1}{II}}}

\defcitealias{Camps2016}{C16}
\defcitealias{Trayford2017}{T17}
\newcommand{\C}{\citetalias{Camps2016}}
\newcommand{\T}{\citetalias{Trayford2017}}

\submitjournal{ApJS}

\shorttitle{UV to submm fluxes for EAGLE galaxies}
\shortauthors{Camps et al.}

\begin{document}

\title{Data release of UV to submm broadband fluxes\\for simulated galaxies from the EAGLE project}

\correspondingauthor{Peter Camps}
\email{peter.camps@ugent.be}
\author[0000-0002-4479-4119]{Peter Camps}
\affiliation{Sterrenkundig Observatorium, Universiteit Gent, Krijgslaan 281, B-9000 Gent, Belgium}
\author{Ana Tr\u cka}
\affiliation{Sterrenkundig Observatorium, Universiteit Gent, Krijgslaan 281, B-9000 Gent, Belgium}
\author{James Trayford}
\affiliation{Institute for Computational Cosmology, Department of Physics, University of Durham, South Road, Durham DH1 3LE, UK}
\author{Maarten Baes}
\affiliation{Sterrenkundig Observatorium, Universiteit Gent, Krijgslaan 281, B-9000 Gent, Belgium}
\author{Tom Theuns}
\affiliation{Institute for Computational Cosmology, Department of Physics, University of Durham, South Road, Durham DH1 3LE, UK}
\author{Robert A. Crain}
\affiliation{Astrophysics Research Institute, Liverpool John Moores University, 146 Brownlow Hill, Liverpool L3 5RF, UK}
\author{Stuart McAlpine}
\affiliation{Institute for Computational Cosmology, Department of Physics, University of Durham, South Road, Durham DH1 3LE, UK}
\author{Matthieu Schaller}
\affiliation{Institute for Computational Cosmology, Department of Physics, University of Durham, South Road, Durham DH1 3LE, UK}
\author{Joop Schaye}
\affiliation{Leiden Observatory, Leiden University, PO Box 9513, NL-2300 RA Leiden, The Netherlands}



\begin{abstract}
We present dust-attenuated and dust emission fluxes for sufficiently resolved galaxies in the EAGLE suite of cosmological hydrodynamical simulations, calculated with the SKIRT radiative transfer code. The post-processing procedure includes specific components for star formation regions, stellar sources, and diffuse dust, and takes into account stochastic heating of dust grains to obtain realistic broad-band fluxes in the wavelength range from ultraviolet to sub-millimeter. The mock survey includes nearly half a million simulated galaxies with stellar masses above $10^{8.5}~\Msun$ across six EAGLE models. About two thirds of these galaxies, residing in 23 redshift bins up to $z=6$, have a sufficiently resolved metallic gas distribution to derive meaningful dust attenuation and emission, with the important caveat that the same dust properties were used at all redshifts. These newly released data complement the already publicly available information about the EAGLE galaxies, which includes intrinsic properties derived by aggregating the properties of the smoothed particles representing matter in the simulation. We further provide an open source framework of Python procedures for post-processing simulated  galaxies with the radiative transfer code SKIRT. The framework allows any third party to calculate synthetic images, SEDs, and broadband fluxes for EAGLE galaxies, taking into account the effects of dust attenuation and emission.
\end{abstract}





\section{Introduction} 

About one third of the stellar light in a typical disk galaxy is reprocessed by interstellar dust before it reaches our telescopes \citep{Soifer1991,Xu1995,Popescu2002,Viaene2016}. The physical processes involved can be probed through multi-wavelength observations in the ultraviolet/optical range (absorption and scattering by dust grains) and in the infrared/sub-millimeter range (thermal emission by dust grains). It has become clear over the years that the star-dust geometry of a galaxy substantially affects its attenuation and emission properties \citep{Byun1994,Corradi1996}, and that even the local, irregular and clumpy structure of the interstellar medium (ISM) has an important global effect \citep{Witt1996,Witt2000,Saftly2015}. Hydrodynamical simulations of galaxy formation routinely attempt to produce this substructure at various scales depending on the resolution of the simulation. Properly comparing the results of these simulations to observations requires solving the complete three-dimensional (3D) radiative transfer (RT) problem to capture the intricate interplay between the simulated galaxy's constituents \citep{Guidi2015, Hayward2015}. In this work we post-process a substantial number of galaxies produced by a recent simulation effort, EAGLE, and we publish the resulting broadband fluxes in a range including ultraviolet (UV), optical, infrared (IR) and sub-millimeter (submm) wavelengths.

The EAGLE project \citep{Schaye2015, Crain2015} consists of a suite of smoothed particle hydrodynamics (SPH) simulations that follow the formation of galaxies and large-scale structure in cosmologically representative volumes of a standard $\Lambda$ cold dark matter universe. EAGLE employs sub-grid recipes for radiative cooling, star formation, stellar mass loss, black hole growth and mergers, and feedback from stars and accreting black holes. While these recipes are calibrated to reproduce the present-day galaxy stellar mass function and galaxy sizes, the simulation results show good agreement with many observables not considered in the calibration \citep[e.g.,][]{Schaye2015, Lagos2015, Bahe2016, Furlong2015, Furlong2017, Trayford2015, Trayford2016, Segers2016, Crain2017}. The EAGLE suite includes a number of independent simulations or ``models" with varying box size and resolution. The public EAGLE database \citep{McAlpine2016,EAGLEteam2017} offers intrinsic properties for all galaxies (subhalos) in these EAGLE models, for 29 simulation snapshots at redshifts ranging from $z=20$ to present-day. The intrinsic galaxy properties were derived by aggregating the properties of the smoothed particles representing the baryonic and dark matter in the simulation. The optical magnitudes listed in the database do not take into account the presence of dust and thus represent an intrinsic aggregation of the stellar sources using a straightforward \citet{Bruzual2003} single-stellar-population model  for each stellar particle.

\citet{Camps2016} and \citet{Trayford2017}, hereafter respectively \C\ and \T, present a procedure to post-process EAGLE galaxies and produce mock observations that do account for the effects of interstellar dust. They extract the relevant information on star formation regions, stellar sources, and the diffuse dust distribution for each galaxy from the respective EAGLE snapshot, and subsequently perform a full 3D RT simulation using the SKIRT code \citep{Baes2011, Camps2015a}. \T\ study optical colors and spectral indices of EAGLE galaxies at redshift $z=0.1$, while \C\ study far-infrared and dust properties of a small set of EAGLE galaxies selected to match a particular subset of the galaxies in the \emph{Herschel} Reference Survey \citep{Boselli2010, Cortese2012}. Comparing the EAGLE simulation results to observations of the local Universe at multiple wavelengths enables the authors to test their post-processing procedure and fine-tune important parameters such as the dust-to-metal ratio.

In this work we apply the post-processing procedure presented by \C\ and \T\ to all EAGLE galaxies with a stellar mass above $10^{8.5}~\Msun$, for all redshifts, in the six most widely studied EAGLE models. We find that for about two thirds of these galaxies, i.e.\ 316\,389 galaxies residing in snapshots up to redshift $z=6$, the post-processing routine produces a sufficiently resolved dust distribution to calculate meaningful dust-attenuated and dust emission fluxes. We publish rest-frame magnitudes and observer-frame fluxes for these galaxies in 50 standard UV--submm wavelength bands as an addition to the public EAGLE database presented by \citet{McAlpine2016}. Publishing these mock observations enables any interested third party to study the dust-related properties of the EAGLE galaxies at all redshifts, and to compare them to observations.

In \autoref{Methods.sec} we describe our methods for post-processing the EAGLE galaxies and for preparing mock observables. We also present the open source framework of Python procedures used for this work, and we indicate how it can be used with minor changes by any third party to calculate synthetic images, integrated spectra (SEDs), and broadband fluxes. In \autoref{PublishedData.sec} we describe the database tables and fields added to the public EAGLE database as a result of this work. In \autoref{Tests.sec} we perform some checks on the published data and show some initial, basic results. Finally, in \autoref{Conclusions.sec} we conclude and provide an outlook to forthcoming work comparing the published fluxes to observations.


\section{Methods}
\label{Methods.sec}

\subsection{Post-processing EAGLE galaxies}
\label{PostProcessingGalaxies.sec}

For a detailed presentation of the EAGLE project (``Evolution and Assembly of GaLaxies and their Environments") we refer to \citet{Schaye2015} and \citet{Crain2015}, and the references therein. In \autoref{PublishedData.sec}, we briefly introduce the six models in the EAGLE suite of simulations for which additional data are being published as part of this work. Here, we just point out a particular characteristic of the EAGLE simulations that is relevant to the RT post-processing procedure employed for this work. Specifically, the EAGLE simulations do not model the cold gas phase in the ISM \citep[see Sect.~4.3 of ][]{Schaye2015}. To prevent artificial fragmentation of star-forming gas, the EAGLE simulations impose a temperature floor, $T_\mathrm{eos}(\rho)$, as a function of the local gas density, $\rho$, corresponding to the polytropic equation of state $\rho\,T_\mathrm{eos}\propto P_\mathrm{eos}\propto\rho^{4/3}$ \citep{Schaye2008}. As a consequence, there are no resolved molecular clouds. Instead, the simulated ISM consists of fairly smoothly distributed, warm gas. Following \C\ and \T, our post-processing procedure addresses the lack of a cold phase by employing a separate sub-grid model for star-forming regions, and by assigning dust to star-forming gas particles regardless of their imposed, unphysical temperature. It remains important, however, to keep this limitation in mind when interpreting our results.

We use the procedure presented in section 2.4 of \C\ to extract galaxies from the EAGLE snapshots and prepare them as RT input models, using the ``standard" parameter values as determined by \C. In summary:
\begin{itemize}
\item We define a galaxy in an EAGLE snapshot as a gravitationally bound substructure in a halo of dark and baryonic matter, as identified by the friends-of-friends and SUBFIND algorithms \citep{Springel2001,Dolag2009b} run on the output of the EAGLE simulations.
\item For each galaxy, we extract the star particles and gas particles within a radius of 30~proper~kpc centered on the galaxy's stellar center of mass. We define a face-on view looking down from the positive net stellar angular momentum vector of the galaxy, an edge-on view observing from an arbitrary direction perpendicular to this vector, and a ``random" view corresponding to the galaxy's original orientation in the simulation volume.
\item From these two particle sets, we move all star particles younger than 100~Myr and all gas particles with a nonzero star formation rate (SFR) into an intermediate set of ``star-forming region'' candidates. All other particles, i.e.\ older star particles and non-star-forming gas particles, are transferred directly to the corresponding two RT input sets.
\item We re-sample the star-forming region candidates into a number of sub-particles with lower masses drawn randomly from a mass distribution function inspired by observations of molecular clouds in the Milky Way, and we assign a random formation time to each sub-particle, assuming a constant SFR over a 100~Myr lifetime.
\item We place the sub-particles that formed less than 10~Myr ago into a third RT input set defining star-forming regions, and we add those that formed more than 10~Myr ago to the input set of star particles, and those that have not yet formed to the set of gas particles.
\item To derive the diffuse dust distribution, we assign a dust mass to all ``cold" gas particles, i.e.\ gas particles with a nonzero SFR \emph{or} with a temperature below $T_\mathrm{max}=8000~\mathrm{K}$, assuming a fixed dust-to-metal fraction $f_\mathrm{dust}=0.3$.
\item To determine the emission spectrum of the stellar sources (other than star-forming regions) in each location, we assign a stellar population SED from the \citet{Bruzual2003} family to each star particle based on its birth mass, metallicity, and age.
\item For the particles in the third input set representing star-forming regions, we follow a special procedure. Following \citet{Jonsson2010}, we assign an appropriate starburst SED from the MAPPINGS III family \citep{Groves2008} to each particle, which models the \HII\ region and the photodissociation region (PDR) surrounding the star-forming core. The SED models both the attenuated starlight and the thermal dust emission emanating from the star-forming region. We calculate the required parameter values from the intrinsic particle properties, with the exception of the time-averaged dust covering fraction of the PDR, which we set to a constant value of $f_\mathrm{PDR}=0.1$.
\item To avoid double counting the dust in the PDR modeled by the MAPPINGS III SEDs, we subtract the implicit PDR dust masses from the diffuse dust distribution surrounding the star-forming region.
\end{itemize}

Given these input sets, we perform RT simulations using the same code as used by \C\ and \T. SKIRT\footnote{SKIRT home page: http://www.skirt.ugent.be\label{skirthome.fn}} is an open source\footnote{SKIRT code repository: https://github.com/skirt \label{skirtrepo.fn}} multi-purpose 3D Monte Carlo dust RT code for astrophysical systems \citep{Baes2011, Camps2015a}. It offers full treatment of absorption and multiple anisotropic scattering by the dust, computes the temperature distribution of the dust and the thermal dust re-emission self-consistently, and supports stochastic heating of dust grains \citep{Camps2015b}. The code handles multiple dust mixtures and arbitrary 3D geometries for radiation sources and dust populations, including grid- or particle-based representations generated by hydrodynamical simulations \citep{Baes2015}. It employs advanced grids for spatial discretization \citep{Saftly2013,Saftly2014} and is fully parallelized using multiple threads and/or multiple processes so that it can run efficiently on a wide range of computing system architectures \citep{Verstocken2017}. 

We use the SKIRT configuration presented in section 2.5 of \C, with some adjustments as noted below. In summary:
\begin{itemize}
\item We discretize the spatial domain using an octree grid that automatically subdivides cells until each cell contains less than a fraction $\delta_\mathrm{max}=3\times10^{-6}$ of the total dust mass in the galaxy, or until 10 subdivisions have been performed. For a domain size corresponding to the 30~kpc radius of our galaxy extraction procedure, the smallest dust cells are thus $2\times30\,\mathrm{kpc}/2^{10}\approx 60\,\mathrm{pc}$ on a side, which offers 5-10 times better resolution than the typical gravitational softening length in the EAGLE simulations.
\item We use the \citet{Zubko2004} dust model to represent the diffuse dust, and (through the MAPPINGS III templates) a similar but not identical dust model for the star-forming regions.
\item We include the effects of stochastically heated dust grains (SHGs) and polycyclic aromatic hydrocarbon molecules (PAHs) in the calculation.
\item We employ a wavelength grid for the RT calculations consisting of 450 wavelength points from 0.02 to 2000 \micron\ laid out on a logarithmic scale, with smaller bin widths in important regions including the PAH emission range and specific emission or absorption features in the employed input spectra.
\item We launch $5\times10^5$ photon packages for each of the $450$ points in the wavelength grid for each of the primary emission and dust emission phases.
\item We place mock detectors along face-on, edge-on, and random viewing angles (see the particle extraction description earlier in the current section) to accumulate spatially integrated fluxes at each wavelength grid point. These detectors are placed at an arbitrary ``local" distance of 20~Mpc.
\end{itemize}

Allowing for the needs of the current work, we adjust the SKIRT configuration used by \C\ as follows:
\begin{itemize}
\item We limit the dust grid domain to an origin-centered cube that just encloses all of the actual dust in the galaxy, rather than always using the full 30 kpc aperture. This improves the spatial resolution in the RT simulations for compact galaxies, which occur more frequently at the higher redshifts considered in this work. 
\item We self-consistently calculate the self-absorption of dust emission by dust. The iteration is considered to converge when the total absorbed dust luminosity is less than one per cent of the total absorbed stellar luminosity, or when the total absorbed dust luminosity has changed by less than three per cent compared to the previous iteration. Dust self-absorption is particularly important for compact, strongly star-forming galaxies because the dust is heated to higher temperatures. As reported in \autoref{ModelVariations.sec}, our tests show that for some EAGLE galaxies the luminosity in submm bands can be underestimated by a factor of 2.5 when ignoring dust self-absorption.
\item We do not produce fully resolved images in the RT simulations for this work. Calculating integrated fluxes is computationally less demanding, and this is an important consideration in view of the large number of EAGLE galaxies to be processed. The lack of a spatially resolved data cube implies that we cannot emulate the observational limitations for the \emph{Herschel} SPIRE 250/350/500 instruments as described by \C. We will see in \autoref{Tests.sec} that this decreases the scatter in the submm color-color relations displayed by the EAGLE galaxies, making the results slightly more ``synthetic". On the other hand, emulating these observational limitations for the submm instruments would have been less meaningful in view of the varying redshifts and the correspondingly large luminosity distances considered in this work.
\end{itemize}

Finally, we process the SEDs detected by the mock instruments in the RT simulation to obtain broadband magnitudes and fluxes:
\begin{itemize}
\item To obtain broadband magnitudes in the galaxy's rest frame for a given viewing angle, we convolve the detected SED with the corresponding response curves and convert the resulting fluxes to absolute AB magnitudes, taking into account the fixed assumed galaxy-detector distance of 20~Mpc.
\item To obtain fluxes in the observer frame, we first shift the detected SED by the galaxy's redshift; then we convolve the shifted SED with the broadband response curves; and finally we scale the broad-band fluxes using
\begin{equation}
f_{\nu,\mathrm{obs}}=(1+z)\,\left(\frac{20~\mathrm{Mpc}}{D_\mathrm{L}}\right)^2\,f_{\nu,\mathrm{shifted}}
\end{equation}
where $z$ is the galaxy's redshift and $D_\mathrm{L}$ the corresponding luminosity distance.
\end{itemize}
With respect to the last item, we determine the luminosity distance from the redshift for each EAGLE snapshot assuming the cosmological parameters used in the EAGLE simulations. Following the suggestions by \citet{Baes2017}, we use the approximation for the luminosity distance presented by \citet{Adachi2012}. We include the calculated luminosity distances in the published data (see \autoref{PublishedData.sec}). For galaxies in redshift zero snapshots, we keep the fluxes at the fixed ``local" distance of 20~Mpc.

\begin{table*}
\caption{The EAGLE models in the public database considered by this work. Columns from left to right: model name; comoving box size; initial baryonic particle mass; the number of galaxies with stellar mass above $10^{8.5}\,\Msun$ for all 29 snapshots; the number of galaxies in this set with $N_\mathrm{dust}>0$ (``some dust") and with $N_\mathrm{dust}>250$ (``resolved dust"), where $N_\mathrm{dust}$ is the number of smoothed (sub-)particles defining the dust content (see \autoref{SelectionGalaxies.sec}).}
\label{models.tab}
\def\arraystretch{1.05} 
\begin{tabular}{l r r r r r}
\hline
EAGLE model  & $L$     & $m_\mathrm{g}$ &  \multicolumn{3}{c}{Number of galaxies with $M_*>10^{8.5}\,\Msun$} \\
                         & (cMpc) & ($\Msun$)           &  All  &  With some dust & With resolved dust \\
\hline
RefL0025N0752            &  25 & $2.26\times10^5$ &    8 279 & 8 096 (97.8\%) & 7 819 (94.4\%) \\
RecalL0025N0752        &  25 & $2.26\times10^5$ &    5 954 & 5 886 (98.9\%) & 5 700 (95.7\%) \\
RefL0025N0376            &  25 & $1.81\times10^6$ &    5 742 & 5 553 (96.7\%) & 3 871 (67.4\%) \\
RefL0050N0752            &  50 & $1.81\times10^6$ &  48 261 & 44 470 (92.1\%) & 31 422 (65.1\%) \\
AGNdT9L0050N0752   &  50 &  $1.81\times10^6$ &  48 278 & 44 601 (92.4\%) & 31 231 (64.7\%) \\
RefL0100N1504            & 100 & $1.81\times10^6$ & 371 728 & 334 717 (90.0\%) & 236 346 (63.6\%) \\
\quad Total                     &       &                               & 488 242 & 443 323 (90.8\%) & 316 389 (64.8\%) \\
\hline
\end{tabular}
\end{table*}

\begin{table*}
\caption{The EAGLE snapshots up to redshift $z=6$. The first three columns list the snapshot number as used in the public database and the corresponding redshift $z$ and luminosity distance $D_\mathrm{L}$. The remaining columns indicate the number of galaxies with stellar mass above $10^{8.5}\,\Msun$ and sufficiently resolved dust ($N_\mathrm{dust}>250$) for each EAGLE model and snapshot.}
\label{snapshots.tab}
\def\arraystretch{1.05} 
\begin{tabular}{r r r  r r r  r r r}
\hline
           &          &                              &  \multicolumn{6}{c}{Number of galaxies with $M_*>10^{8.5}\,\Msun$ and resolved dust ($N_\mathrm{dust}>250$)} \\
Snap   & $z$   &  $D_\mathrm{L}$  &  Ref & Recal & Ref & Ref & AGNdT9 & Ref \\
Num    &          &  (Mpc)                   &  L0025N0752 & L0025N0752 & L0025N0376 & L0050N0752 & L0050N0752 & L0100N1504 \\
\hline
  28 &  0.00 & $2.00\times 10^1$ &    486 &    369 &    140 &  1 048 &  1 011 &  7 101 \\
 27 &  0.10 & $4.79\times 10^2$ &    527 &    384 &    155 &  1 150 &  1 138 &  8 072 \\
 26 &  0.18 & $9.16\times 10^2$ &    544 &    390 &    162 &  1 237 &  1 228 &  8 744 \\
 25 &  0.27 & $1.43\times 10^3$ &    561 &    393 &    184 &  1 341 &  1 359 &  9 600 \\
 24 &  0.37 & $2.02\times 10^3$ &    564 &    395 &    198 &  1 465 &  1 444 & 10 428 \\
 23 &  0.50 & $2.94\times 10^3$ &    558 &    388 &    217 &  1 655 &  1 652 & 11 846 \\
 22 &  0.62 & $3.75\times 10^3$ &    547 &    381 &    238 &  1 743 &  1 773 & 12 782 \\
 21 &  0.74 & $4.66\times 10^3$ &    524 &    372 &    246 &  1 924 &  1 920 & 14 086 \\
 20 &  0.87 & $5.69\times 10^3$ &    506 &    362 &    255 &  2 054 &  2 026 & 15 269 \\
 19 &  1.00 & $6.83\times 10^3$ &    492 &    347 &    275 &  2 148 &  2 124 & 16 143 \\
 18 &  1.26 & $9.04\times 10^3$ &    450 &    308 &    263 &  2 295 &  2 252 & 17 001 \\
 17 &  1.49 & $1.11\times 10^4$ &    390 &    286 &    267 &  2 327 &  2 259 & 17 228 \\
 16 &  1.74 & $1.34\times 10^4$ &    337 &    256 &    249 &  2 196 &  2 209 & 16 561 \\
 15 &  2.01 & $1.61\times 10^4$ &    298 &    233 &    227 &  2 003 &  2 009 & 15 445 \\
 14 &  2.24 & $1.83\times 10^4$ &    278 &    214 &    219 &  1 842 &  1 814 & 14 128 \\
 13 &  2.48 & $2.07\times 10^4$ &    249 &    193 &    188 &  1 639 &  1 628 & 12 517 \\
 12 &  3.02 & $2.63\times 10^4$ &    181 &    148 &    146 &  1 162 &  1 177 &  9 368 \\
 11 &  3.53 & $3.17\times 10^4$ &    128 &    108 &    100 &    793 &    802 &  6 783 \\
 10 &  3.98 & $3.66\times 10^4$ &     92 &     80 &     69 &    549 &    562 &  4 916 \\
  9 &  4.49 & $4.21\times 10^4$ &     55 &     48 &     43 &    371 &    364 &  3 399 \\
  8 &  5.04 & $4.82\times 10^4$ &     28 &     26 &     18 &    223 &    222 &  2 139 \\
  7 &  5.49 & $5.33\times 10^4$ &     16 &     14 &      8 &    143 &    141 &  1 412 \\
  6 &  5.97 & $5.88\times 10^4$ &      7 &      4 &      4 &     78 &     79 &    901 \\
\hline
\end{tabular}
\end{table*}

\subsection{Uncertainties}
\label{Uncertainties.sec}

Although the presented procedure has been validated by \C\ and \T, it is important to note the sources of uncertainties in the results and the related caveats. We consider three categories of uncertainty, ignoring any limitations of the EAGLE simulation methods themselves (because evaluating those limitations is why we produce mock observations to begin with). Firstly, EAGLE galaxies are represented in the generated snapshots with a limited resolution. The stellar and/or ISM distribution in some galaxies might not be sufficiently resolved to allow meaningful 3D RT results. This is further explored in \autoref{PublishedData.sec} and  \autoref{Tests.sec}.

Secondly, the discretization of the RT problem introduces interpolation errors and noise:
\begin{itemize}
\item Re-sampling the star-forming region candidates into a number of sub-particles is a randomized process; a different sequence of (pseudo-)random numbers will result in a galaxy with slightly different properties.
\item Approximating the spatial domain through a dust grid and representing the wavelength range by a number of discrete bins causes interpolation errors.
\item The Monte Carlo technique introduces Poisson noise due to the finite number of photon packages.
\end{itemize}
From the convergence tests performed by \C\ and \T\ and some additional tests conducted for this work, we conclude that the combined uncertainty on the calculated broadband magnitudes caused by these numerical limitations is $\pm0.05$ mag.

Thirdly, there are issues introduced by the choices made during the design of the procedure. Most notably:
\begin{itemize}
\item The calculated fluxes depend on the particular viewing angle selected by the procedure. The galactic plane, and thus the face-on position, is ill-defined for irregular galaxies, and thus may vary with subtle changes in the procedure. The edge-on viewing angle can be chosen from any of the $2\pi$ directions perpendicular to the face-on direction. While many disk galaxies are fairly axisymmetric, for some less regular galaxies the dust-attenuated flux can vary substantially from one edge-on sight line to another.
\item The \citet{Zubko2004} dust model (with absorption coefficient at $350~\micron$ of $\kappa_{350}=0.330~\mathrm{m}^2\,\mathrm{kg}^{-1}$ and power-law index $\beta=2$) is used for all galaxies, regardless of redshift or galaxy type, while its grain composition and size distribution have been fine-tuned for interstellar dust in the Milky Way.
\item Similarly, the procedure uses fixed values for the dust-to-metal ratio ($f_\mathrm{dust}=0.3$) and PDR covering factor ($f_\mathrm{PDR}=0.1$), while these calibrated values were obtained by \C\ and \T\ (in the context of post-processing the EAGLE simulations) for a set of galaxies in the local Universe, i.e.\ $z\leq0.1$.
\end{itemize}

In \autoref{ModelVariations.sec} we evaluate the effects of some variations to our post-processing procedure that seem particularly relevant. Interested parties can further explore these and other model adjustments for a selection of EAGLE galaxies using the open-source code framework discussed in \autoref{pythonframework.sec}.

\subsection{The Python framework}
\label{pythonframework.sec}

Performing the presented procedure for nearly half a million EAGLE galaxies cannot be done without appropriate automation. While most of the processing time is consumed by the actual RT simulation in the SKIRT code, there is a fair amount of pre- and post-processing and overall data management involved as well. We implemented all of these extra functions in the programming language Python, adding them to the open source Python Toolkit for SKIRT (PTS). The PTS code can be downloaded from a public repository (see \autoref{skirtrepo.fn}) and the PTS documentation is hosted on the SKIRT web site (see \autoref{skirthome.fn}). Please refer to the topic on post-processing EAGLE galaxies in the online PTS User Guide. Here we limit the discussion to a brief summary of the PTS functionality related to this work.

Our EAGLE Python framework is designed to run on a large computing system with multiple nodes governed by a job scheduling system. We assume that all computing nodes have access to a common file system that contains all input and output data files. The overall post-processing workflow is managed through a simple SQLite database that includes a record for each requested RT simulation run. This ``run" record specifies the EAGLE galaxy to be processed and the SKIRT configuration to be used for the RT simulation, in addition to some fields that keep track of its current workflow state. The Python procedures allow a user to insert new run records in the database, support the scheduling of jobs on the system to move these runs through the various workflow stages (extract, simulate, observe), and finally enable the collection of the results into a single data set. The workflow stages have been separated so that the scheduled jobs can be adjusted to the specific resource requirements for each stage (e.g.\ the extraction procedure runs in a single thread, while a SKIRT simulation can use multiple parallel threads or even multiple nodes).

While we believe the data published as part of this work will form a sufficient basis for many science projects, in some cases it may be meaningful to re-process a selection of EAGLE galaxies with an updated version of our Python procedures. Because both our Python code and the EAGLE snapshot particle data are publicly available \citep{EAGLEteam2017}, any interested party can undertake such a project. Implementation of the required adjustments will often be straightforward or even trivial. For example:
\begin{itemize}
\item Produce a full 3D data cube (integral field unit) with a resolved image for every point in the wavelength grid.
\item Include more viewing angles.
\item Use another dust model (material properties, grain size distribution).
\item Vary the dust-related parameters in the procedure, such as the dust-to-metal ratio.
\item Vary the SED templates assigned to stellar populations.
\item Adjust the treatment of star-forming regions.
\end{itemize}
Also, our galaxy extraction module can be adapted to process the output of hydrodynamical simulations other than EAGLE without affecting the remainder of the Python framework. The PTS documentation provides further guidance for making these and other changes.


\section{Published data}
\label{PublishedData.sec}

\subsection{Resolution criteria for selecting EAGLE galaxies}
\label{SelectionGalaxies.sec}

\autoref{models.tab} lists the six EAGLE models considered in this work, with the respective box sizes and mass resolutions. For a more detailed description of the various models, see tables 2 and 3 and the accompanying text in \citet{Schaye2015}. The fourth column of \autoref{models.tab} indicates the number of galaxies with a stellar mass above $10^{8.5}\,\Msun$ for each model, accumulated over all 29 snapshots. This stellar mass cutoff matches the set of galaxies for which the public EAGLE database includes optical magnitudes without dust attenuation. 

We performed the procedure presented in \autoref{PostProcessingGalaxies.sec} on all 488 242 galaxies with stellar mass above $10^{8.5}\,\Msun$. The average runtime per galaxy was nearly 43 node-minutes, for a total runtime of 39.6 node-years. Given that we used 16-core nodes (with a Sandy Bridge architecture), this is equivalent to more than 630 years of serial processing. For all simulations combined, SKIRT launched and traced more than $3.7\times10^{14}$ photon packages.

The mock observables resulting from a RT simulation are meaningful only if the input distributions for both the stellar sources and the body of dust are spatially resolved to an acceptable level. We use the number of relevant SPH (sub-)particles as a measure for the numerical resolution of each of these density distributions:
\begin{gather}
N_\mathrm{star} = \max(N_\mathrm{*},N_\mathrm{SFR}) \label{Nstar.eq} \\
N_\mathrm{dust} = \max(N_\mathrm{coldgas},N_\mathrm{SFR}) \label{Ndust.eq}
\end{gather}
where $N_\mathrm{*}$, $N_\mathrm{SFR}$ and $N_\mathrm{coldgas}$, respectively, indicate the number of (sub-)particles in the sets representing young and evolved stars, star-forming regions, and cold gas particles. As described in \autoref{PostProcessingGalaxies.sec}, these sets may contain original SPH particles extracted from the EAGLE snapshot and/or resampled sub-particles replacing star-forming region candidates. Because the star-forming regions contribute both to the optical and submm fluxes, they are counted towards both $N_\mathrm{star}$ and $N_\mathrm{dust}$. We use the maximum operator rather than addition to obtain a slightly more stringent criterion, considering that the subsampled particles are not fully independent of each other. It turns out that, with these definitions, $N_\mathrm{star}>N_\mathrm{dust}$ for all processed galaxies. Also, in the context of RT, getting the stellar distribution exactly right is arguably less important than properly resolving the dust distribution. We can thus focus on $N_\mathrm{dust}$ as a measure of numerical resolution for our purposes.

\begin{figure}
\centering
\includegraphics[width=\columnwidth]{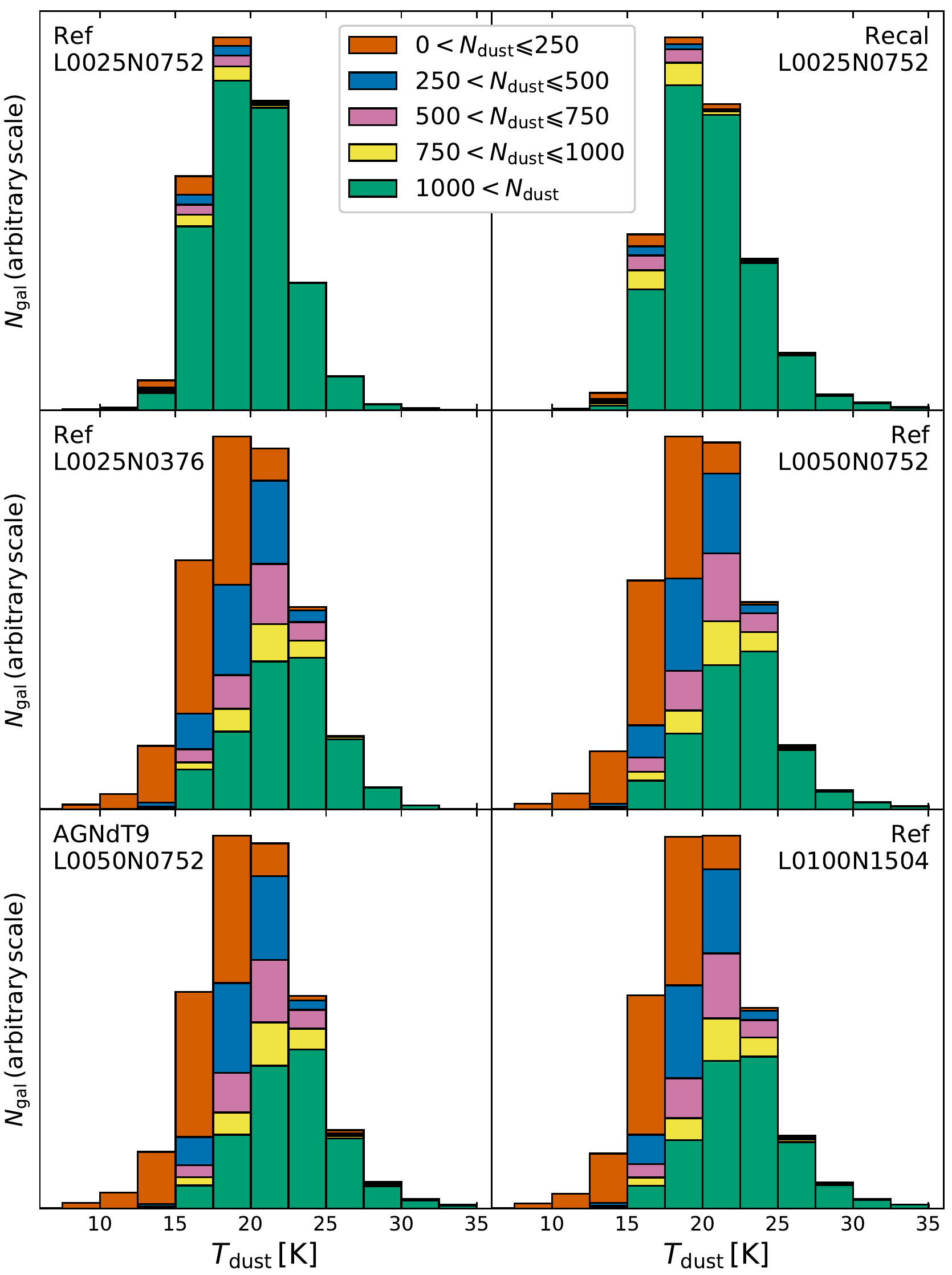}
\caption{Distribution of the representative dust temperature $T_\mathrm{dust}$ for the galaxies within each EAGLE model. The vertical scale is adjusted for each panel to fit the highest histogram bar. The upper two panels show the high-resolution models, the lower four panels the regular-resolution models (see \autoref{models.tab}). The overlapping histograms include subsets of galaxies with increasing numbers of particles representing dust, $N_\mathrm{dust}$.}
\label{DustTempHistograms.fig}
\end{figure}

\begin{figure*}
\centering
\includegraphics[width=\textwidth]{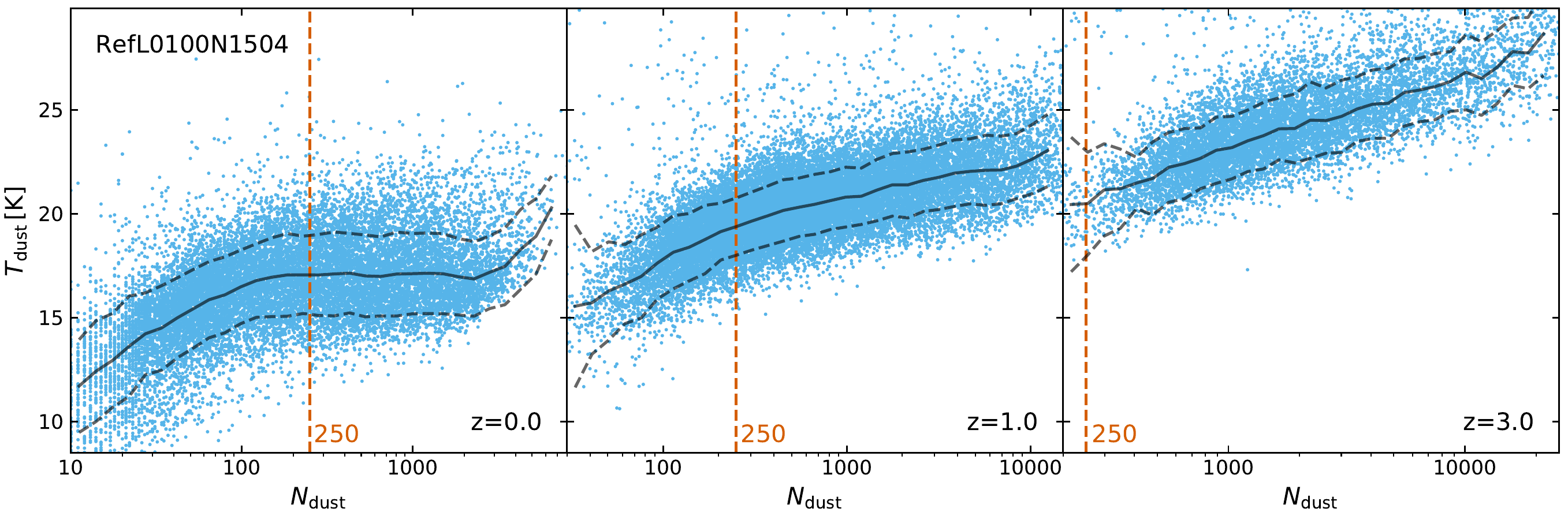}
\caption{The representative dust temperature, $T_\mathrm{dust}$, as a function of the number of particles representing dust, $N_\mathrm{dust}$, for the galaxies in the RefL0100N1504 model at three different redshifts; from left to right $z=0,1,3$. The solid curve traces the median temperature; the dashed curves indicate the standard deviation. The red dashed vertical line indicates the cutoff value of $N_\mathrm{dust}$. Galaxies to the right of this line are considered to be sufficiently resolved.}
\label{CountCorrelation.fig}
\end{figure*}

To help evaluate the quality of the calculated fluxes as a function of input resolution, we estimate the total dust mass $M_\mathrm{dust}$ and the representative dust temperature $T_\mathrm{dust}$ for each galaxy from the fluxes in the continuum dust emission range. Specifically, we fit a modified black body (MBB) curve to the \emph{Herschel} PACS 160 and SPIRE 250, 350 and 500 bands, converting the rest-frame absolute magnitudes in the database to rest-frame fluxes at an arbitrary ``local" distance. We use a MBB with power-law index $\beta=2$ and absorption coefficient $\kappa_{350}=0.330~\mathrm{m}^2\,\mathrm{kg}^{-1}$, matching the dust model in our post-processing procedure, and free parameters $T_\mathrm{dust}$ and $M_\mathrm{dust}$. The fit employs a least-squares Levenberg-Marquardt optimization algorithm, allowing for three times more uncertainty on the outer data points (160 and 500 \micron) than on the inner data points (250 and 350 \micron).

\autoref{DustTempHistograms.fig} shows the distribution of $T_\mathrm{dust}$ so obtained for the galaxies within each EAGLE model. The upper two panels show the high-resolution models, the lower four panels the regular-resolution models (see third column of \autoref{models.tab}). The overlapping histograms include subsets of galaxies with increasing numbers of particles representing dust, $N_\mathrm{dust}$. There is also a fraction of galaxies that have no particles representing dust, i.e.\ $N_\mathrm{dust}=0$. For these galaxies, there is no submm flux and the dust fitting algorithm cannot be performed, so they are omitted from this figure. However, this is the case for less than 10 per cent of the galaxies for all models (see fifth column of \autoref{models.tab}).

It is easily seen from the histograms in \autoref{DustTempHistograms.fig} that many of the galaxies with $0<N_\mathrm{dust}\leq250$ (shown in blue) have an ch{unrealistically} low dust temperature of $T_\mathrm{dust}<15~\mathrm{K}$. While this is especially evident for the regular-resolution models, the same trend is present for the higher-resolution models, although they include a much smaller fraction of such galaxies. The artificially low temperatures can be understood by realizing that the dust density distribution in these galaxies is numerically gravely under-sampled, and that the dust may be improperly placed relative to the primary radiation sources. More generally, the histograms for consecutive $N_\mathrm{dust}$ bins show that the temperature distribution becomes more symmetrical when including only galaxies with larger values of $N_\mathrm{dust}$, and that the average temperature increases. The latter trend is at least in part explained by the fact that galaxies with a high SFR, and thus higher average dust temperatures, are likely to include many sub-particles representing star formation regions.

\begin{figure}
\centering
\includegraphics[width=0.88\columnwidth]{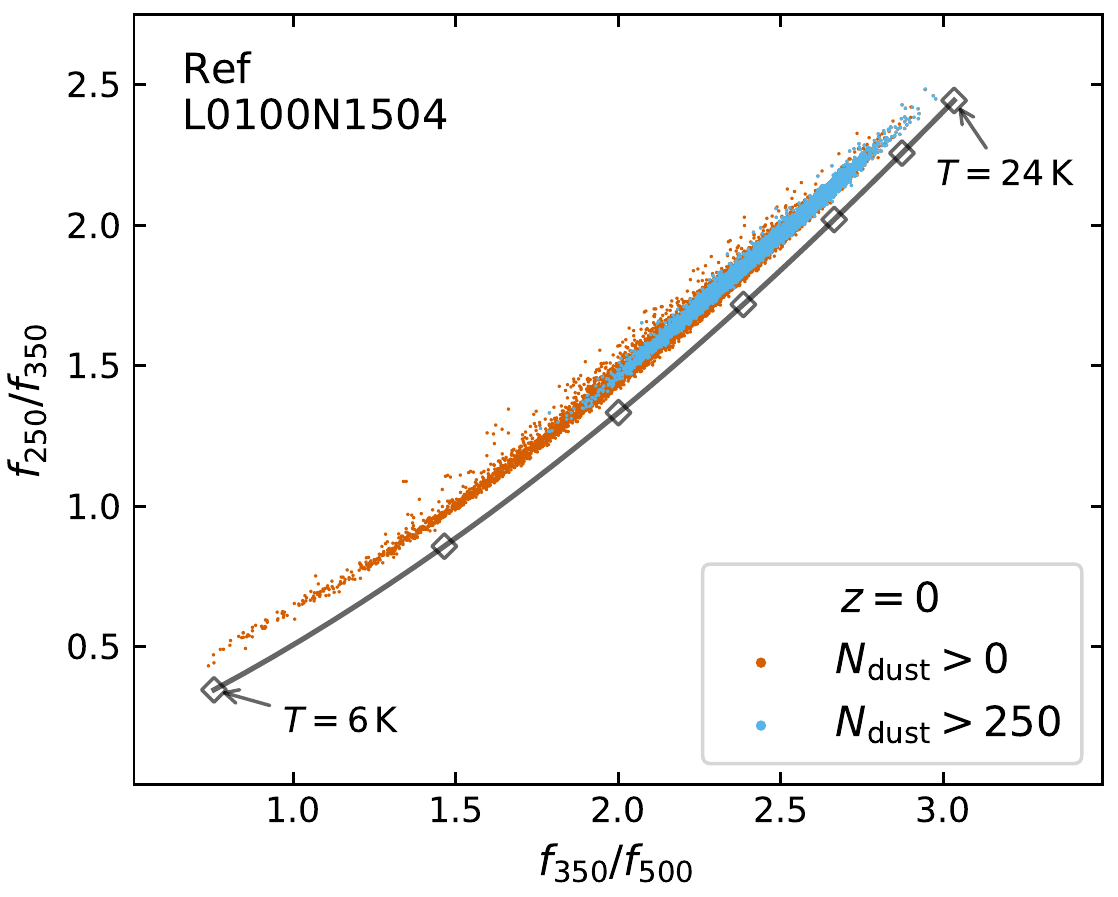}
\caption{\emph{Herschel} SPIRE color-color relation $f_{250}/f_{350}$ versus $f_{350}/f_{500}$ for the EAGLE galaxies at redshift zero in the RefL0100N1504 model. This corresponds to figure 11 of \C. The red dots represent all galaxies in the snapshot that have at least some dust; the cyan dots represent the subset of galaxies that satisfy our numerical resolution criterion. The solid curve traces a modified black body (MBB) with $\beta=2$ for temperatures ranging from 6~K to 24~K; the diamonds are spaced by 3~K.}
\label{SubmmColorsRedshiftZero.fig}
\end{figure}

In \autoref{CountCorrelation.fig} this effect is illustrated in more detail for the galaxies in three snapshots of the RefL0100N1504 model, or equivalently, in three different redshift bins. For the redshift zero galaxies (left panel), the median temperature is essentially constant in the range $250<N_\mathrm{dust}\leq2500$. The steep rise for  $N_\mathrm{dust}\leq250$ is hard to explain on physical grounds and is probably caused by the poor numerical resolution. The rise beyond $N_\mathrm{dust}>2500$ is probably caused by the correlation with star formation rate mentioned earlier. For the galaxies at redshift $z=1$ (middle panel), the situation seems to be similar, although the knees in the median temperature curve are less prominent. For much higher redshifts (right panel), most galaxies have $N_\mathrm{dust}>250$, which again may be related to their increased star formation rate and dust content.

A related effect of the numerical resolution is illustrated in \autoref{SubmmColorsRedshiftZero.fig}, which shows a \emph{Herschel} SPIRE color-color relation for present-day EAGLE galaxies in the RefL0100N1504 model, corresponding to figure 11 of \C. The SPIRE submm fluxes characterize the downwards slope of the dust continuum emission, and thus are sensitive to the cold dust content. Smaller flux ratios $f_{250}/f_{350}$ and $f_{350}/f_{500}$ indicate a flatter slope of the dust emission curve and thus a larger contribution from colder dust. This is illustrated in the figure by the solid curve, which traces the flux ratio relation for the emission of a MBB with $\beta=2$ (the value assumed by the dust model used in this work). The red dots in our figure represent all galaxies in the snapshot that have at least some dust; the cyan dots represent the subset of galaxies with $N_\mathrm{dust}>250$. It is again obvious that many of the galaxies with $N_\mathrm{dust}\leq250$ have unrealistically low temperatures.

Although it is impossible to unambiguously derive a precise criterion for galaxies that are sufficiently resolved, based on these statistics, we opt to publish dust-attenuated and dust emission fluxes and magnitudes for galaxies with $N_\mathrm{dust}>250$. The last column of \autoref{models.tab} lists the number of galaxies that satisfy this criterion for each model. This amounts to roughly 64 per cent of the total number of galaxies for the regular-resolution models, and roughly 95 per cent for the high-resolution models. \autoref{snapshots.tab} provides an overview of the number of galaxies that satisfy the criterion per snapshot, or equivalently, per redshift bin. It also lists the luminosity distance $D_\mathrm{L}$ used to scale the observer-frame fluxes in the database.

\subsection{Selection bias}

\begin{figure*}
\centering
\includegraphics[width=\textwidth]{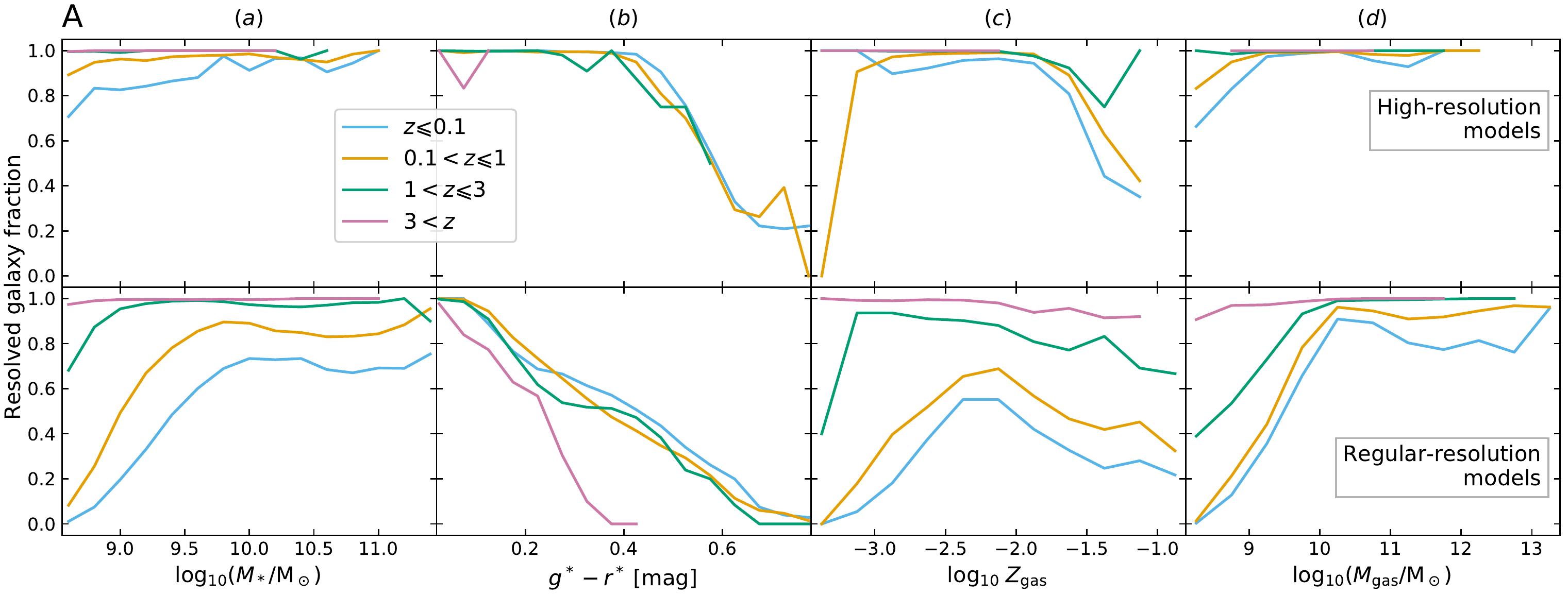}\vspace{3mm}
\includegraphics[width=\textwidth]{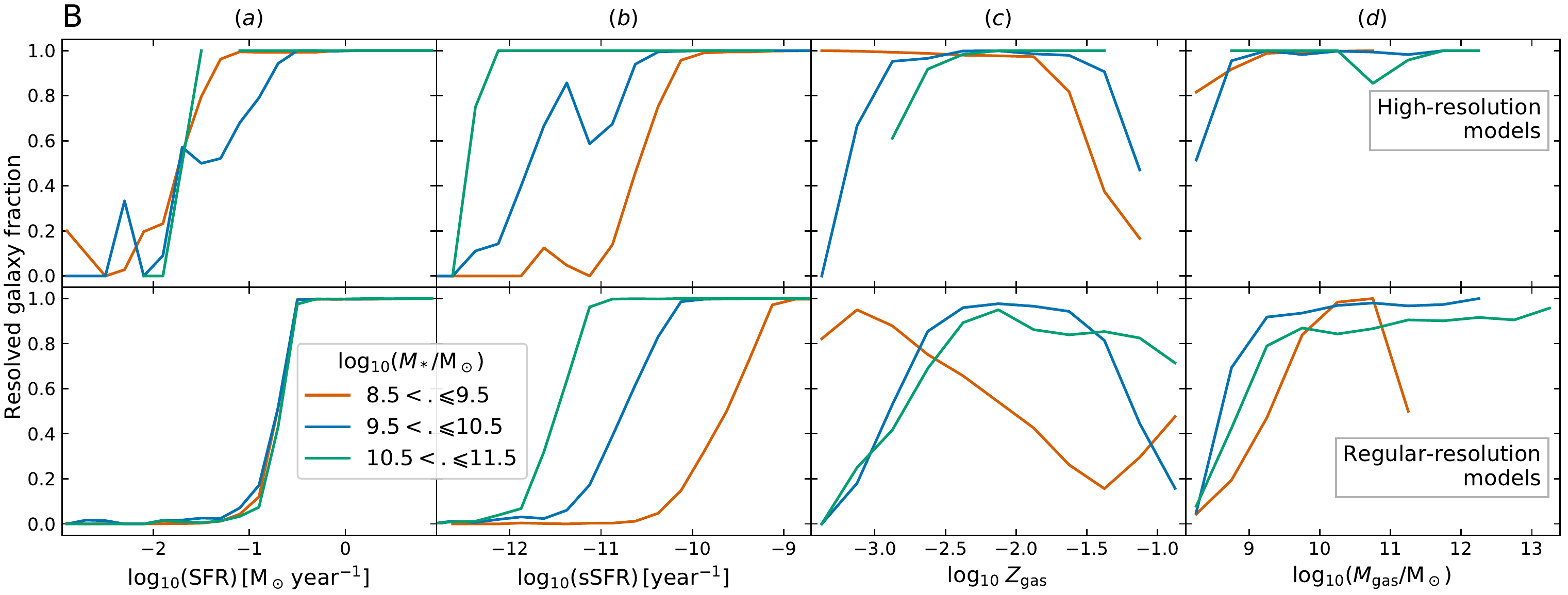}
\caption{The fraction of sufficiently resolved ($N_\mathrm{dust}>250$) EAGLE galaxies as a function of various intrinsic galaxy properties, i.e.\ properties directly derived from the EAGLE simulation output without RT post-processing. The top half of the figure (part A) shows resolved galaxy fractions as a function of stellar mass, intrinsic $g\!*-r*$ color (ignoring any effects of dust), gas metallicity (as a plain metal fraction, not normalized to solar metallicity), and gas mass (including both star-forming and non-star-forming gas). All panels in part A use the four redshift bins listed in the figure legend. The bottom half of the figure (part B) similarly shows resolved galaxy fractions as a function of star formation rate (SFR), specific SFR, gas metallicity, and gas mass. The panels in part B use the three stellar mass bins listed in the figure legend. In each figure part, the top row shows aggregate fractions for the high-resolution EAGLE models (RefL0025N0752 and RecalL0025N0752), and the bottom row shows aggregate fractions for the regular-resolution models (RefL0025N0376, RefL0050N0752, AGNdT9L0050N0752, and RefL0100N1504).}
\label{FigResolvedBin.fig}
\end{figure*}

Excluding EAGLE galaxies that have insufficient numerical resolution for modeling dust, using a threshold on the number of dust-related input particles as described in the previous section, unavoidably introduces a selection bias. \autoref{FigResolvedBin.fig} quantifies the bias introduced by our selection as a function of various intrinsic galaxy properties, i.e.\ properties directly derived from the EAGLE simulation output without RT post-processing. Specifically, the panels in the figure show the fraction of sufficiently resolved ($N_\mathrm{dust}>250$) EAGLE galaxies as a function of stellar mass, dust-free $g\!*-r*$ color, SFR, specific SFR, gas metallicity, and gas mass. The top half of the figure (part A) shows fractions for four redshift bins with borders at $z=0.1$, 1, and 3, so that the first bin corresponds to the local Universe. The bottom half of the figure (part B) uses three stellar mass bins centered respectively at $M_*=10^9$, $10^{10}$, and $10^{11}~\Msun$. In each figure part, the top row shows aggregate fractions for the high-resolution EAGLE models, and the bottom row shows aggregate fractions for the regular-resolution models.

As an overall trend, the high-resolution models have a higher resolved galaxy fraction for low stellar masses (\autoref{FigResolvedBin.fig}A column \emph{a}) and low star formation rates (\autoref{FigResolvedBin.fig}B columns \emph{a} and \emph{b}) than the regular-resolution models. This is obviously because the high-resolution models use a larger number of particles to represent a given dust mass, so that the galaxies in these models stay above the threshold more often. Furthermore, within a particular model, galaxies with lower (specific) star formation rates are much more often excluded (\autoref{FigResolvedBin.fig}B columns \emph{a} and \emph{b}); the precise threshold depends on the resolution of the model and, for the specific SFR, on the stellar mass, as can be seen in column \emph{b} of \autoref{FigResolvedBin.fig}B. Similarly, red galaxies are much more often excluded than blue galaxies (\autoref{FigResolvedBin.fig}A column \emph{b}). In other words, quiescent ellipticals are more likely to be excluded than actively star-forming spirals, because the former contain much less dust and thus are more likely to fall below a threshold based on the number of dust-related input particles.

As a function of average gas metallicity (column \emph{c} of \autoref{FigResolvedBin.fig}, A and B), for most models the resolved galaxy fraction remains fairly constant over the range $0.002 < Z_\mathrm{gas}< 0.03$. A notable exception is the drop in the resolved fraction for the lowest mass bin in the regular-resolution models (\autoref{FigResolvedBin.fig}B column \emph{c}), caused again by the fact that these lower-mass galaxies do not contain a sufficient number of dust-related particles to make the threshold. 

Comparing the curves for the various redshift bins in \autoref{FigResolvedBin.fig}A reveals a number of relevant points as well. When plotted as a function of stellar mass (\autoref{FigResolvedBin.fig}A column \emph{a}), the resolved fraction increases significantly with redshift, especially for the  regular-resolution models. For redshifts up to $z\approx2$ this is plausible because star formation increases with redshift in this range \citep{Madau2014}, leading to a higher number of dust-related particles. For even higher redshifts ($z>3$), the number of EAGLE galaxies above the stellar mass threshold of $10^{8.5}~\Msun$ decreases rapidly (see \autoref{snapshots.tab}), and the galaxies that do make it above the mass threshold are likely to be rather active as a result of recent mergers. When plotted as a function of intrinsic color (\autoref{FigResolvedBin.fig}A column \emph{b}), the resolved fraction is fairly constant with redshift up to $z\approx3$. It decreases significantly for higher redshifts ($z>3$), especially for red galaxies ($g\!*-r*>0.2$ mag). This can again be traced to the fact that the high-redshift galaxies above the mass threshold are likely to be active.

The evolution of the resolved galaxy fraction as a function of gas mass (column \emph{d} of \autoref{FigResolvedBin.fig}, A and B) is qualitatively similar to the evolution as a function of stellar mass (\autoref{FigResolvedBin.fig}A column \emph{a}). This is not surprising because of the correlation between gas and stellar mass, even if the relation has significant scatter. The curve for the lowest stellar mass bin in column \emph{d} of \autoref{FigResolvedBin.fig}B is cut off at $M_\mathrm{gas}\approx 10^{11}~\Msun$ because the bin contains no galaxies with that much gas.

\begin{table*}
\caption{The database tables and fields published as a result of this work, where \emph{Model} is replaced by each of the EAGLE model names listed in \autoref{models.tab}, and \emph{Band} by each of the broadband field names listed in \autoref{bands.tab}.}
\label{tables.tab}
\def\arraystretch{1.03}
\setlength{\tabcolsep}{2pt}
\begin{tabular}{l l}
\hline
Table/field name & Description \\
\hline
\emph{Model}\_ParticleCounts  &  Numerical resolution measures for the galaxy; for all galaxies \\
\quad GalaxyID & \quad Galaxy identifier (unique within each model) \\
\quad Count\_Star & \quad Number of (sub-)particles $N_\mathrm{star}$ representing the galaxy's stellar sources (see \autoref{Nstar.eq}) \\
\quad Count\_Dust & \quad Number of (sub-)particles $N_\mathrm{dust}$ representing the galaxy's body of dust (see \autoref{Ndust.eq}) \\
\emph{Model}\_DustFit  &  Results of fitting a modified black body to restframe submm fluxes; only for galaxies with $N_\mathrm{dust}>0$ \\
\quad GalaxyID & \quad Galaxy identifier (unique within each model) \\
\quad Temp\_Dust & \quad Estimated representative dust temperature $T_\mathrm{dust}$ in K\\
\quad Mass\_Dust & \quad Estimated dust mass $M_\mathrm{dust}$ in $\Msun$ \\
\emph{Model}\_DustyMagnitudes  &  Rest-frame absolute magnitudes; only for galaxies with $N_\mathrm{dust}>250$ \\
\quad GalaxyID & \quad Galaxy identifier (unique within each model) \\
\quad \emph{Band} & \quad Absolute AB magnitude in the rest frame of the galaxy \\
\emph{Model}\_DustyFluxes  &  Observer-frame fluxes; only for galaxies with $N_\mathrm{dust}>250$ \\
\quad GalaxyID & \quad Galaxy identifier (unique within each model) \\
\quad \emph{Band} & \quad Flux in Jy observed in a frame taking into account the galaxy's redshift and luminosity distance \\
Snapshots  &  Redshift and luminosity distance for each snapshot (i.e.\ the first three columns of \autoref{snapshots.tab}) \\
\quad SnapNum & \quad Snapshot number \\
\quad Redshift & \quad Redshift \\
\quad LumDistance & \quad Luminosity distance $D_\mathrm{L}$ in Mpc \\
\hline
\end{tabular}
\end{table*}

\begin{table}
\caption{The instruments or filters for which mock broad-band observed fluxes and absolute AB magnitudes are calculated and included in the public database. The first column in each table specifies the database field name, the second column indicates the corresponding pivot wavelength. The table on the left lists UV/optical bands, for which there are actually three fields in the database. The field name has an optional suffix (not shown in the table) indicating the viewing angle: ``\texttt{\_e}" for edge-on, ``\texttt{\_f}" for face-on, and no suffix for random view. The table on the right lists IR/submm bands, for which there is only a single field because emission in these bands is essentially isotropic.}
\label{bands.tab}
\def\arraystretch{1.05}
\setlength{\tabcolsep}{1pt}
\begin{minipage}{\columnwidth}
\begin{tabular}{l r}
\hline
Field name & $\lambda_\mathrm{pivot}$ (\micron) \\
\hline
GALEX\_FUV  & 0.1535  \\
GALEX\_NUV  & 0.2301  \\
SDSS\_u     & 0.3557  \\
SDSS\_g     & 0.4702  \\
SDSS\_r      & 0.6176  \\
SDSS\_i      & 0.7490  \\
SDSS\_z     & 0.8947  \\
TwoMASS\_J  & 1.239  \\
TwoMASS\_H  & 1.649  \\
TwoMASS\_Ks & 2.164  \\
UKIDSS\_Z  & 0.8826  \\
UKIDSS\_Y  & 1.031  \\
UKIDSS\_J  & 1.250  \\
UKIDSS\_H  & 1.635  \\
UKIDSS\_K  & 2.206  \\
Johnson\_U   & 0.3525  \\
Johnson\_B   & 0.4417  \\
Johnson\_V   & 0.5525  \\
Johnson\_R   & 0.6899  \\
Johnson\_I    & 0.8739  \\
Johnson\_J   & 1.243  \\
\hline
\end{tabular}%
\begin{tabular}{l r}
\hline
Field name & $\lambda_\mathrm{pivot}$ (\micron) \\
\hline
Johnson\_M  & 5.012  \\
WISE\_W1    & 3.390  \\
WISE\_W2    & 4.641  \\
WISE\_W3    & 12.57  \\
WISE\_W4    & 22.31  \\
IRAS\_12    & 11.41  \\
IRAS\_25    & 23.61  \\
IRAS\_60    & 60.41  \\
IRAS\_100    & 101.1  \\
IRAC\_I1    & 3.551  \\
IRAC\_I2    & 4.496  \\
IRAC\_I3    & 5.724  \\
IRAC\_I4    & 7.884  \\
MIPS\_24   & \tablenotemark{*}23.76 \\
MIPS\_70   & \tablenotemark{*}71.99 \\
MIPS\_160 & \tablenotemark{*}156.4 \\
PACS\_70    & 70.77  \\
PACS\_100   & 100.8  \\
PACS\_160   & 161.9  \\
SPIRE\_250 & 252.5  \\
SPIRE\_350 & 354.3  \\
SPIRE\_500 & 515.4  \\
SCUBA2\_450 & 449.3  \\
SCUBA2\_850 & 853.8  \\
ALMA\_10 & 349.9  \\
ALMA\_9 & 456.2  \\
ALMA\_8 & 689.6  \\
ALMA\_7 & 937.9  \\
ALMA\_6 & 1244  \\
\hline
\end{tabular}
\end{minipage}
\tablenotetext{*}{In table 4 of \C\ the \emph{Spitzer} MIPS instruments are inadvertently listed as bolometers. Properly classifying these instruments as photon counters results in slightly adjusted pivot wavelengths.} 
\end{table}

\subsection{Database tables and fields}

As a result of this work, the public EAGLE database\footnote{Public EAGLE database: http://www.eaglesim.org/database.php} is extended with several tables as listed in \autoref{tables.tab}. Most tables are repeated for each EAGLE model, indicated by including the model name in the table name. The only exception is the \texttt{Snapshots} table, which contains information that is valid for all models. Except for the \texttt{Snapshots} table, the first field in each of the new tables is the \texttt{GalaxyID}, an integer number that uniquely identifies a galaxy within a particular model. The same identifier is also used in the previously published part of the EAGLE database \citep[see][]{McAlpine2016}. In other words, this field can be used to join any of the tables in the public EAGLE database.

The \texttt{ParticleCounts} tables contain the values of $N_\mathrm{star}$ and $N_\mathrm{dust}$ as defined in \autoref{SelectionGalaxies.sec} for all processed galaxies, i.e.\ for all galaxies in the EAGLE database with $M_*>10^{8.5}\,\Msun$. This information is provided as a measure of the numerical resolution of the RT simulation input for a galaxy, allowing users of the database to select galaxies above a certain resolution limit. 

The \texttt{DustFit} tables provide the values of $T_\mathrm{dust}$ and $M_\mathrm{dust}$, estimated as presented in \autoref{SelectionGalaxies.sec}, for all galaxies with $N_\mathrm{dust}>0$. Galaxies that have no particles representing dust are omitted from these tables because the MBB fitting procedure cannot be performed without fluxes in the submm range. The data in the \texttt{DustFit} tables can easily be obtained from the observables in the \texttt{DustyMagnitudes} tables (except for galaxies with $N_\mathrm{dust}\leq250$ which are omitted from those tables; see next paragraph). It is provided merely as a convenience so that the estimated dust mass and temperature can be used in database queries.

For galaxies with $N_\mathrm{dust}>250$, the \texttt{DustyMagnitudes} tables contain absolute AB magnitudes in the rest frame of the galaxy, and the \texttt{DustyFluxes} tables contain fluxes expressed in Jy and observed in a present-day frame taking into account the galaxy's redshift. These quantities are directly derived from the RT simulation output as described at the end of \autoref{PostProcessingGalaxies.sec}. Each table contains fields for the broad-bands listed in \autoref{bands.tab}. For the UV/optical bands (listed in the lefthand portion of \autoref{bands.tab}), there are actually three fields in the database. The field name has an optional suffix indicating the viewing angle: ``\texttt{\_e}" for edge-on, ``\texttt{\_f}" for face-on, and no suffix for random view. For the IR/submm bands (listed in the righthand portion of \autoref{bands.tab}), there is only a single field because emission in these bands is essentially isotropic. As discussed in \autoref{Uncertainties.sec}, we estimate the combined uncertainty on the calculated broadband magnitudes and fluxes resulting from numerical noise in the post-processing procedure to be $\pm0.05$ mag.

The \texttt{Snapshots} table includes the snapshot number and corresponding redshift and luminosity distance for each of the 29 snapshots in the EAGLE models. This information is also listed in the first three columns of \autoref{snapshots.tab}. It is provided as part of the database so that it can be used in database queries.

\begin{figure}
\centering
\begin{framed}
\begin{minipage}{0.9\columnwidth}
\begin{verbatim}
SELECT
    ape.SFR as SFR,
    flx.GALEX_NUV_e as NUV_e,
    flx.GALEX_NUV_f as NUV_f,
    flx.MIPS_24 as M24
FROM
    RefL0100N1504_SubHalo as gal,
    RefL0100N1504_Aperture as ape,
    RefL0100N1504_ParticleCounts as cnt,
    RefL0100N1504_DustyFluxes as flx
WHERE
    gal.SnapNum = 28 and
    gal.Spurious = 0 and
    ape.ApertureSize = 30 and
    cnt.Count_Dust > 250 and
    gal.GalaxyID = ape.GalaxyID and
    gal.GalaxyID = cnt.GalaxyID and
    gal.GalaxyID = flx.GalaxyID
\end{verbatim}
\end{minipage}
\end{framed}
\caption{Example SQL query on the extended public EAGLE database. The query returns the intrinsic star formation rate, edge-on and face-on NUV fluxes, and 24 \micron\ fluxes for all sufficiently resolved present-day galaxies in the database.}
\label{QueryExample.fig}
\end{figure}

\citet{McAlpine2016} describe how to access and query the public EAGLE database. \autoref{QueryExample.fig} presents an example SQL query accessing the extended database to retrieve the intrinsic star formation rate, edge-on and face-on NUV fluxes, and 24 \micron\ fluxes for all sufficiently resolved present-day galaxies. This information is plotted in \autoref{StarFormationRate.fig}, which is discussed in \autoref{Tests.sec}.


\section{Checks and examples}
\label{Tests.sec}

\begin{figure}
\centering
\includegraphics[width=0.85\columnwidth]{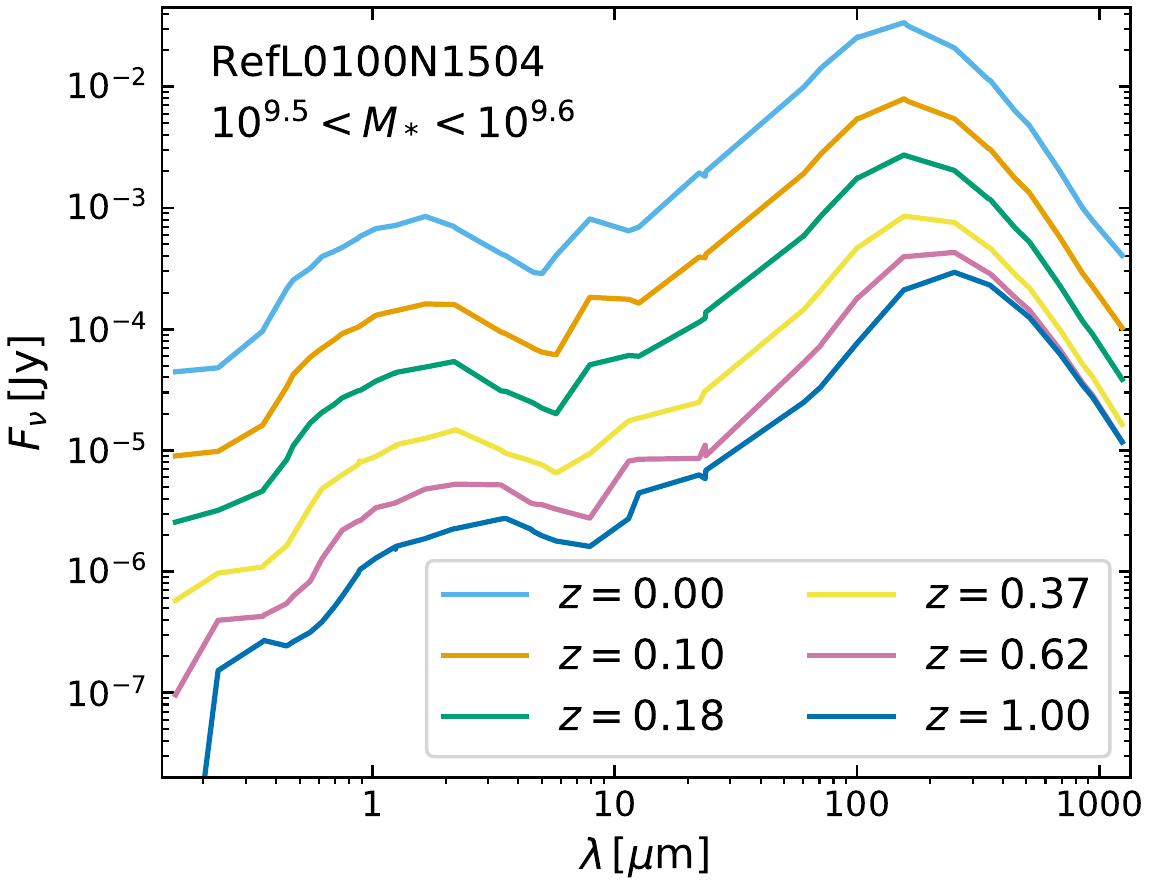}
\caption{Stacked SEDs for RefL0100N1504 galaxies in a narrow stellar mass range of $10^{9.5}<M_*<10^{9.6}$ for redshifts bins (snapshots) from $z=0$ to $z=1$. Each SED is obtained by averaging the fluxes for the more than 500 galaxies in the corresponding mass/redshift bin, and plotting this average flux at the pivot wavelength for each band in the database (see \autoref{bands.tab}). The fluxes are scaled to the luminosity distance corresponding to each bin. For display purposes, the fluxes for $z=0$ are scaled to a distance of 200~Mpc instead of the 20~Mpc assumed in the database.}
\label{StackedSEDs.fig}
\end{figure}

\begin{figure}
\centering
\includegraphics[width=0.7\columnwidth]{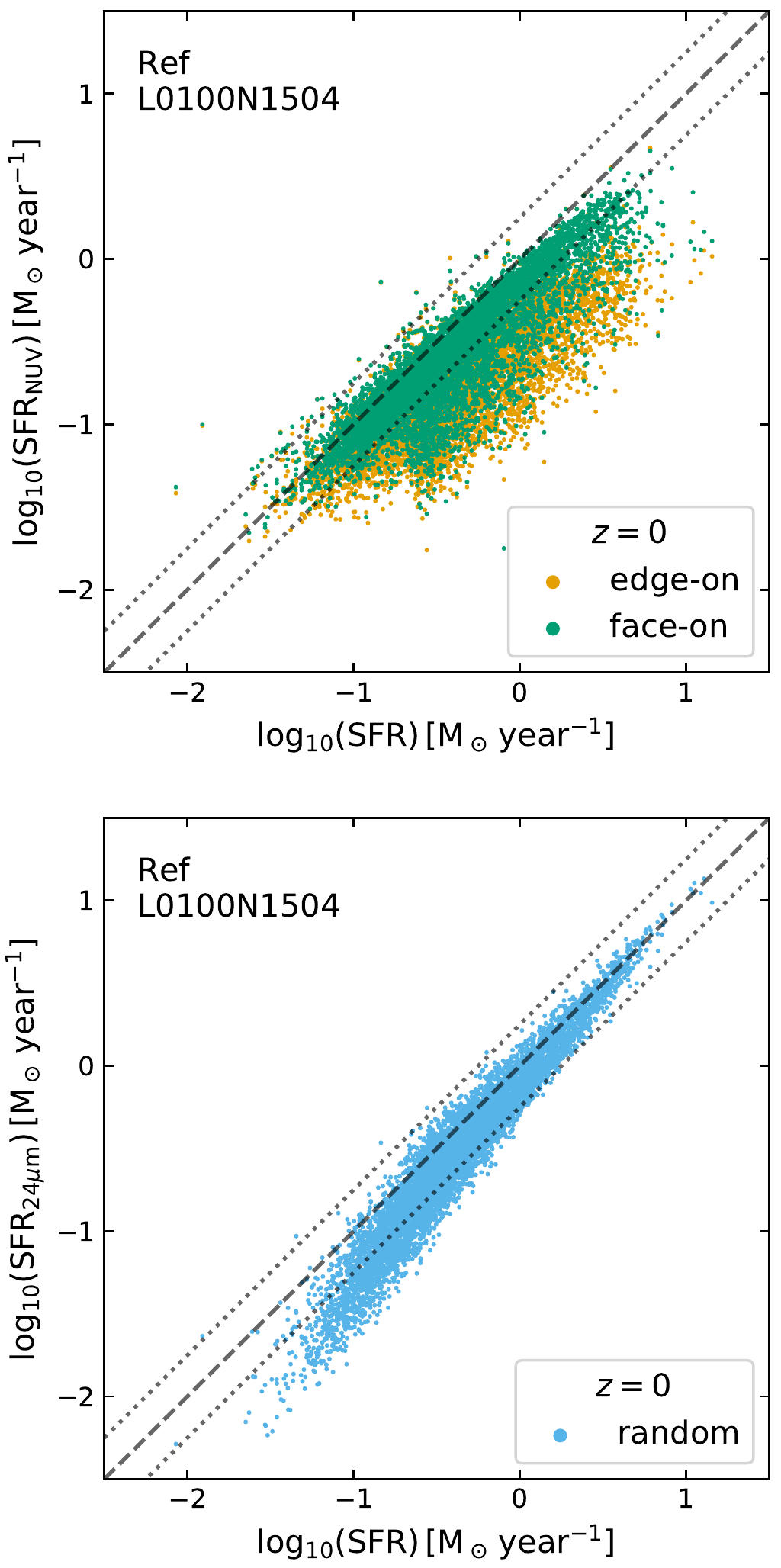}
  \caption{Comparison of two star-formation-rate (SFR) indicators to the intrinsic SFR for the 7100 redshift zero RefL0100N1504 galaxies that satisfy our resolution criterion. This corresponds to figure 10 in \C, where only 282 galaxies were shown. The dashed diagonal in each panel indicates the one-to-one relation; the dotted lines indicate $\pm$0.25 dex offsets. Upper panel: SFR based on GALEX\_NUV flux \citep{Hao2011,Murphy2011}. Lower panel: SFR based on MIPS\_24 flux \citep{Rieke2009}. }
\label{StarFormationRate.fig}
\end{figure}

\begin{figure}
\centering
\includegraphics[width=0.85\columnwidth]{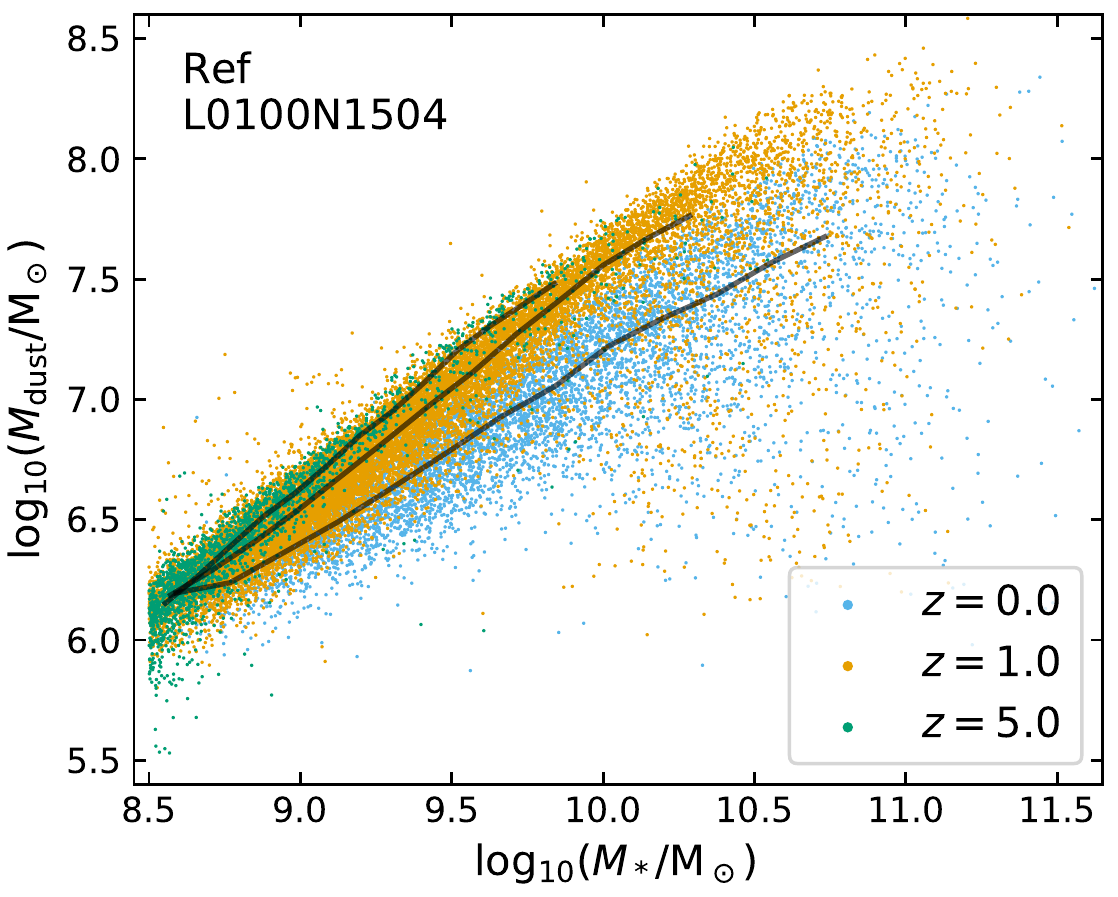}
\caption{Estimated dust mass as a function of intrinsic stellar mass for the RefL0100N1504 galaxies at the redshifts $z=0$ (cyan), $z=1$ (orange), and $z=5$ (green), over-plotted in that order. The solid black lines indicate the running median for each of the three populations.}
\label{DustStellarMass.fig}
\end{figure}

\begin{figure*}
\centering
\includegraphics[width=\textwidth]{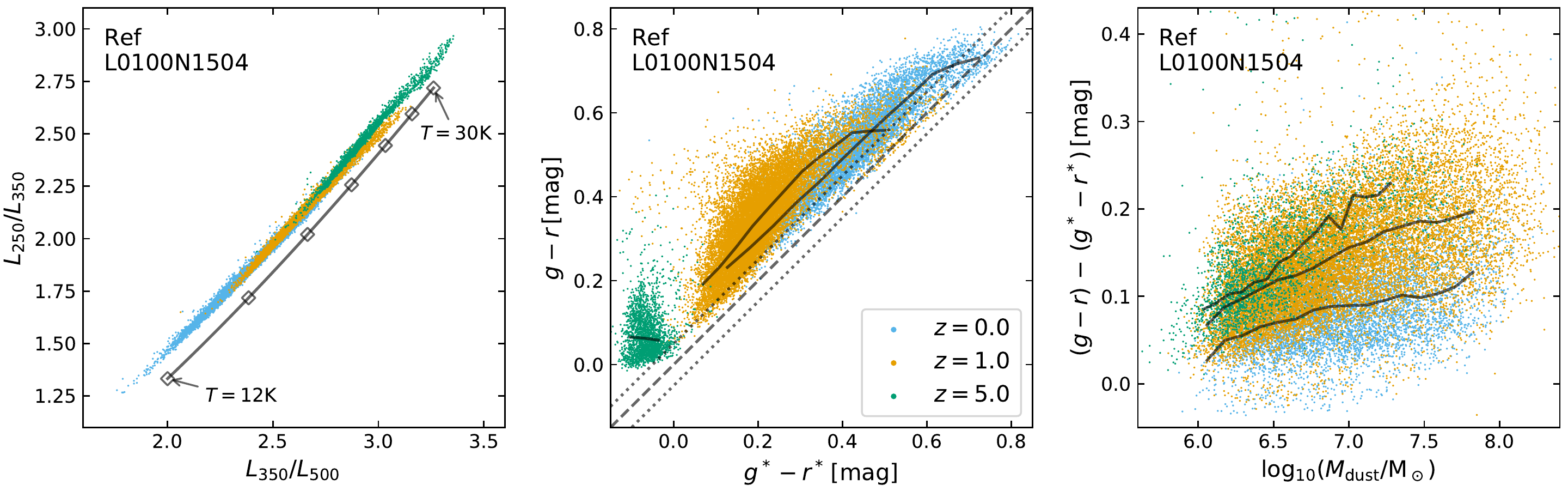}
\caption{Rest-frame scaling relations based on the absolute magnitudes for the RefL0100N1504 galaxies at the redshifts $z=0$ (cyan), $z=1$ (orange), and $z=5$ (green), over-plotted in that order.
\emph{Left}: Submm color-color relation $L_{250}/L_{350}$ versus $L_{350}/l_{500}$ corresponding to \autoref{SubmmColorsRedshiftZero.fig} but excluding the galaxies that do not satisfy our resolution criterion; the cyan dots are the same in both figures. The solid curve traces a modified black body (MBB) with $\beta=2$ for temperatures ranging from 12~K to 30~K; the diamonds are spaced by 3~K.
\emph{Middle}: Dust-affected $g-r$ color for a random orientation (SDSS\_g $-$ SDSS\_r) versus the intrinsic, dust-free $g\!*-r*$ color (g\_nodust $-$ r\_nodust). This corresponds to figure 5 of \T. The dashed diagonal indicates the one-to-one relation; the dotted lines indicate the $\pm 0.05$ mag numerical uncertainty on the calculated magnitudes; the solid black lines indicate the running median for each of the three populations.
\emph{Right}: The difference between the dust-attenuated and dust-free colors of the previous panel, i.e.\ the amount of reddening, versus the estimated dust mass. The solid black lines indicate the running median for each of the three populations.}
\label{ColorsMultiRedshift.fig}
\end{figure*}

We performed several checks on the data described in \autoref{PublishedData.sec} and published as part of this work. For example, we reproduced many of the figures in \C\ and \T\ using a larger number of galaxies and/or including higher redshifts. Rather than listing a repetitive series of plots that attempt to cover all aspects of the data, we present here a small selection of plots that illustrate key points or offer relevant insights. All plots in this section are for the EAGLE reference model RefL0100N1504 and include only galaxies with $N_\mathrm{dust}>250$.

\subsection{Basic tests and scaling relations}

As a first basic test, \autoref{StackedSEDs.fig} shows stacked SEDs for galaxies in a narrow stellar mass range, for redshift bins from $z=0$ to $z=1$, using averages of the fluxes for the bands stored in the database and the pivot wavelength for each band (see \autoref{bands.tab}). As expected, the SED shape shifts to longer wavelengths with increasing redshift, and the fluxes scale down as a result of the increasing luminosity distance. Because each of the fluxes has been obtained from the convolution with a broadband filter, narrow spectral features are smoothed over. Specifically, the fine structure of the infrared emission by SHGs and PAHs is no longer visible, although the simulated spectra from which the broadband fluxes are calculated do resolve these features. The small discontinuities in the SEDs around wavelength $\lambda\approx 23~\micron$ are caused by the overlapping WISE\_W4, IRAS\_25 and MIPS\_24 bands. Variations in the precise position and form of the corresponding filters cause the convolution to pick up different portions of the dust emission spectral features, resulting in small but noticeable differences in the broadband fluxes plotted at nearby pivot wavelengths.

\autoref{StarFormationRate.fig} shows two star-formation-rate (SFR) indicators, calculated using the fluxes in the extended EAGLE database following \citet{Hao2011} and \citet{Murphy2011} for NUV and \citet{Rieke2009} for 24~\micron, compared to the intrinsic SFR already provided in the existing EAGLE database. This corresponds to figure 10 in \C; however, we show all present-day galaxies in the model that satisfy our resolution criterion as opposed to a very limited sample. Note that many of the outliers in figure 10 of \C\ do not satisfy our resolution criterion (i.e.\ they have $N_\mathrm{dust}\leq250$), so that they are not shown in \autoref{StarFormationRate.fig}. Other than this, the results for our larger sample confirm the analysis provided by \C. At the short wavelengths used by the GALEX NUV indicator (our upper panel), the edge-on fluxes (orange) suffer significantly more from dust extinction than the face-on fluxes (green), and thus yield a correspondingly lower inferred SFR. However, even the indicator based on face-on fluxes slightly underestimates the SFR for most galaxies. The indicator based on the \emph{Spitzer} MIPS 24~\micron\ flux (lower panel of \autoref{StarFormationRate.fig}) typically underestimates the SFR of our galaxies. \C\ attribute this at least in part to limitations in the EAGLE simulations (such as the lack of a cold ISM phase) and our post-processing procedure (such as assuming isotropically emitting star-forming regions) that cause some fraction of the diffuse dust in the simulated galaxies to be heated insufficiently, resulting in a 24~\micron\ flux lower than observed. 

\autoref{DustStellarMass.fig} shows the estimated dust mass stored in the extended EAGLE database (and determined as described in \autoref{SelectionGalaxies.sec}) as a function of intrinsic stellar mass already provided in the existing EAGLE database, for the RefL0100N1504 galaxies at the three redshifts $z=0$ (cyan), $z=1$ (orange), and $z=5$ (green). Comparing this figure to observations reported by \citet{Bourne2012} for local galaxies ($z\leq0.35$) and to those reported by \citet{Santini2014} for higher redshifts ($z\leq2.5$), we conclude that these dust masses are within the observed range. The dust mass shows a clear correlation with stellar mass, as expected \citep{Bourne2012}, although with substantial scatter. The scatter increases for the most massive systems ($M_* > 10^{10}\,\mathrm{M}_\odot$) which include elliptical galaxies containing little or no dust \citep{diSeregoAlighieri2013}. Recall that our resolution criterion may cause less massive systems with low dust content to be excluded, slightly biasing the plotted selection. At higher redshifts there are fewer massive systems, and these galaxies contain more dust for the same stellar mass, also as expected \citep{Bourne2012, Santini2014, daCunha2015}. Note that we could replace the intrinsic stellar mass in this plot by a stellar mass indicator derived from observed fluxes. However, as shown in figure 9 of \C, this would most likely introduce just a systematic offset with very limited scatter.

\autoref{ColorsMultiRedshift.fig} presents three scaling relations based on the absolute rest-frame magnitudes stored in the extended EAGLE database, for the same selection of galaxies as in the previous figure. The leftmost panel shows the submm color-color relation corresponding to \autoref{SubmmColorsRedshiftZero.fig} and to figure 11 of \C, but excluding the galaxies that do not satisfy our resolution criterion. The submm fluxes for the EAGLE galaxies shown here follow a tight temperature relation with even less scatter than the EAGLE galaxies presented in figure 11 of \C. Specifically, there are no outliers to the right of the MBB temperature curve. As discussed by \C, these outliers were caused by the simulated observational limitations built into the procedure employed by \C. Because we do not impose such observational limitations in the procedure used for this work, as described in \autoref{PostProcessingGalaxies.sec}, our galaxies stay on the underlying, tight relation. It is also evident from this panel in \autoref{ColorsMultiRedshift.fig} that the overall dust temperature in an EAGLE galaxy generally increases substantially at higher redshifts, with temperatures up to 30~K at $z=5$.

The middle panel of \autoref{ColorsMultiRedshift.fig} relates the dust-affected $g-r$ color stored in the extended EAGLE database to the intrinsic $g^*-r^*$ color of the stellar emission already published in the existing database. The amount of reddening caused by dust extinction is indicated by the vertical distance between a galaxy's representation in the chart and the diagonal one-to-one relation. This corresponds to figure 5 of \T, omitting the inclination information but including higher redshifts. As discussed in \T, intrinsically red ($g^*-r^* > 0.6$) galaxies follow the one-to-one relation closely with little offset, whereas intrinsically blue ($g^*-r^* < 0.4$) galaxies are offset to redder colors and show a large scatter. This trend continues for higher redshifts, with galaxies that are intrinsically a lot bluer, corresponding to the increased star formation rates and more compact dust geometries at those redshifts. A small number of galaxies lie marginally below the one-to-one relation; \T\ attribute this mostly to numerical uncertainties on the calculated magnitudes, although starlight scattered into the line of sight by dust grains might lead to negative attenuation in some cases.

The rightmost panel of \autoref{ColorsMultiRedshift.fig} shows the amount of reddening (i.e.\ the difference between the dust-affected and dust-free colors of the previous panel) versus the dust mass estimated by fitting a MBB to the submm fluxes as described in \autoref{SelectionGalaxies.sec}. Within the population for each redshift, there is a clear trend showing increased reddening for larger inferred dust masses, as expected. The relation has substantial scatter, illustrating the effect of the intrinsic stellar spectrum and the relative stellar/dust geometry on the overall reddening. At the same time, for constant dust mass, the average reddening increases substantially for higher redshifts. This can be understood by noting that higher-redshift galaxies are smaller \citep{vanderWel2014,Furlong2015}, so that the stellar radiation along a particular line of sight encounters more dust (for the same total dust mass) and thus experiences more extinction. This effect is strengthened by the clumpy structure of the dust enveloping star-forming regions, which tend to be more numerous in higher-redshift galaxies.

\subsection{$K$-band dust emission}

\begin{figure}
\centering
\includegraphics[width=0.99\columnwidth]{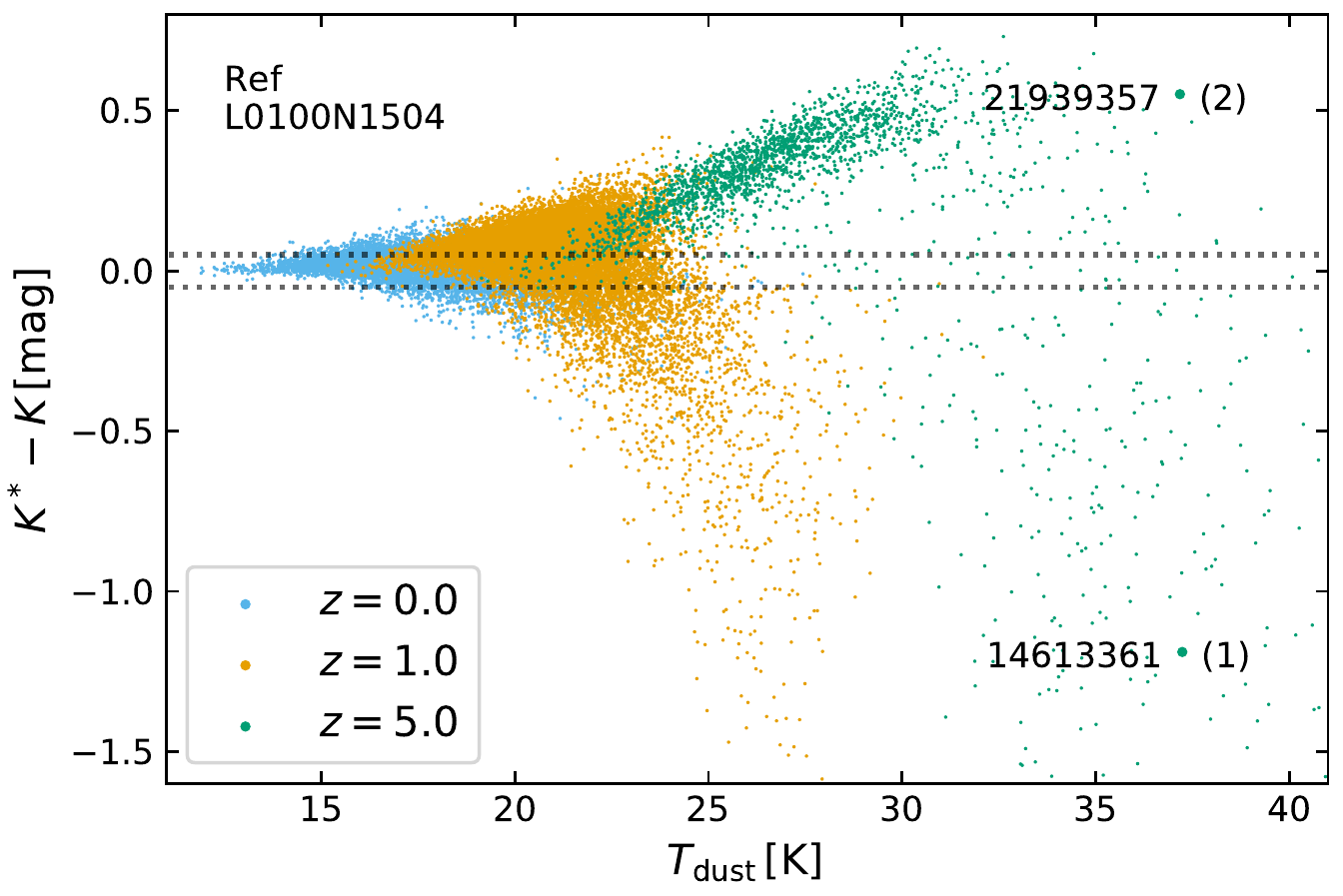}
\caption{The difference between the intrinsic, dust-free $K^*$-band magnitude (K\_nodust) and the dust-affected $K$-band magnitude (UKIDSS\_K), both in the galaxy's restframe, versus the estimated representative dust temperature, for the RefL0100N1504 galaxies at the redshifts $z=0$ (cyan), $z=1$ (orange), and $z=5$ (green), over-plotted in that order. The horizontal dotted lines indicate the $\pm 0.05$ mag numerical uncertainty on the calculated magnitudes. The labeled green dots indicate specific galaxies (with given GalaxyID) that are further discussed in the text and for which properties are provided in \autoref{twogalaxyprops.tab}.}
\label{KbandDiscrepancy.fig}
\end{figure}

\begin{table*}
\caption{Properties of the two EAGLE galaxies labeled in \autoref{KbandDiscrepancy.fig}. Both galaxies reside in the RefL0100N1504 model at redshift $z=5$. In addition to the GalaxyID, columns from left to right list (a) the intrinsic stellar mass, (b) the intrinsic specific star formation rate, (c) the estimated dust mass, (d) the dust to stellar mass ratio, (e) the radius containing 99 per cent of the dust mass, (f) a measure for the average dust surface density, (g) the estimated representative dust temperature, (h) the absolute $K$-band magnitude, (i) the dust emission contribution to the $K$-band luminosity, (j) the attenuated stellar emission contribution to the $K$-band luminosity, and (k) the ratio of the intrinsic dust-free stellar $K$-band luminosity over the observed luminosity. Magnitudes and luminosities are specified in the rest frame for the random viewing angle. The dust radius and the various luminosity contributions are determined from extra calculations; this information is not stored in the public EAGLE database.}
\label{twogalaxyprops.tab}
\def\arraystretch{1.1} \setlength{\tabcolsep}{4pt}
\begin{tabular}{l r r r r r r r r r r r}
\hline
& $_{(a)}$ & $_{(b)}$ & $_{(c)}$ & $_{(d)}$ & $_{(e)}$ & $_{(f)}$ & $_{(g)}$ & $_{(h)}$ & $_{(i)}$ & $_{(j)}$ & $_{(k)}$ \\
GalaxyID & $M_*$ & $\displaystyle\frac{\dot{M_*}}{M_*}$ & $M_\mathrm{dust}$ & $\displaystyle\frac{M_\mathrm{dust}}{M_*}$ & $R_\mathrm{dust}$ 
   & $\displaystyle\frac{M_\mathrm{dust}}{R_\mathrm{dust}^2}$ & $T_\mathrm{dust}$ & $\mathrm{M}_K$ 
   & $\displaystyle\frac{L_{K,\mathrm{dust}}}{L_K}$ & $\displaystyle\frac{L_{K,\mathrm{stars}}}{L_K}$
   & $\displaystyle\frac{L_{K,\mathrm{free}}}{L_K}$ \\
& ($\Msun$) & ($\mathrm{year}^{-1}$)& ($\Msun$) & &  ($\mathrm{kpc}$) & ($\Msun\mathrm{pc}^{-2}$) & ($\mathrm{K}$)  & (mag) \\
\hline
14613361\,(1) & $1.1\times10^{10}$ & $4.4\times10^{-9}$ & $4.9\times10^7$ & $4.5\times10^{-3}$ & 2.9 &5.8& 37.2 & -21.5 & 0.32 & 0.68 & 3.38 \\
21939357\,(2) & $6.3\times10^8$ & $1.3\times10^{-8}$ & $2.5\times10^6$ & $4.0\times10^{-3}$ & 8.5 &0.034& 37.2 & -20.4 & 0.49 & 0.51 & 0.65 \\
\hline
\end{tabular}
\end{table*}

For the same selection of galaxies used in the previous two figures, we now investigate whether dust emission contributes significantly in the (rest-frame) $K$ band, which has a pivot wavelength of about 2.2~\micron\ in the near-IR. \autoref{KbandDiscrepancy.fig} shows the difference between the dust-free $K$-band magnitude already provided in the existing EAGLE database and the dust-affected $K$-band magnitude stored in the extended database, both in the galaxy's restframe, as a function of the estimated representative dust temperature. The vertical axis thus shows the combined effect of dust attenuation and dust emission in the $K$-band. Dust attenuation causes a galaxy to move down, while dust emission causes a galaxy to move up. Galaxies positioned above the zero line (or rather above the 0.05 mag numerical uncertainty) most likely feature a nonzero contribution from dust emission, although there might be some negative attenuation caused by scattering. A significant dust emission contribution becomes more likely with increasing dust temperature because the hotter dust emits at shorter wavelengths, and in general it happens for a substantial fraction of our simulated galaxies with $T_\mathrm{dust}>20$~K. This includes some of the present-day galaxies, a large portion of the galaxies at $z=1$, and most of the galaxies at $z=5$, at least in part because the dust temperature increases with redshift.

The $K$-band emission for galaxies positioned below the zero line in \autoref{KbandDiscrepancy.fig} may still include a relevant contribution from dust emission that is compensated by extinction of the stellar radiation. Unfortunately we cannot disentangle these contributions based on the information stored in the EAGLE database. To shed some light on the matter, we reran the RT process for a few dozen $z=5$ galaxies with $T_\mathrm{dust}>35$~K, this time recording the various contributions separately in the simulation output. The GalaxyID-labeled dots in \autoref{KbandDiscrepancy.fig} represent two extreme galaxies, handpicked for illustrative purposes. \autoref{twogalaxyprops.tab} lists additional properties for these galaxies, extracted in part from the information in the (extended) public EAGLE database, and in part from our extra RT simulations. In the text below we refer to these galaxies through the first digit of their GalaxyID (1 for 14613361 and 2 for 21939357). 
 
Judging from their respective positions in \autoref{KbandDiscrepancy.fig}, galaxy 2 must have substantial $K$-band dust emission, and galaxy 1 must have substantial dust extinction, but we cannot say much about its dust emission. Evaluating the results of our extra simulations, it turns out that in \emph{both} galaxies dust emission represents one third or more of the $K$-band luminosity (column \emph{i} of \autoref{twogalaxyprops.tab}). At the same time, galaxy 1 features strong dust extinction (column \emph{k}), more than compensating for the dust emission contribution, so that it ends up in its low position in \autoref{KbandDiscrepancy.fig}. To understand why this is happening, let us look at the respective galaxy properties in \autoref{twogalaxyprops.tab}. Both galaxies have a similar dust to stellar mass ratio (column \emph{d}) and a comparable specific star formation rate (column \emph{b}), making it plausible that the dust is heated to a similar average temperature (column \emph{g}). However, galaxy 1 is much more massive than galaxy 2 (column \emph{a}), and at the same time it is much more compact (column \emph{e}). The average dust surface density in galaxy 1 is 170 times higher than that in galaxy 2 (column \emph{f}). As a result, the dust in galaxy 1 blocks a lot more stellar radiation along each particular line of sight, explaining the extreme extinction (column \emph{k}).

Our study of the $K$-band results illustrates a number of important points. At least for the simulated EAGLE galaxies, dust emission can contribute significantly to the rest-frame $K$-band luminosity, especially at higher redshifts, and even for some present-day galaxies. Given that our procedure tends to underestimate the dust temperatures (see \C\ and the discussion of \autoref{StarFormationRate.fig} earlier in this section), we can surmise that this should also be the case for observed galaxies \citep[also see, e.g.,][]{Hunt2002}. At the same time, this shows that we need to perform a full panchromatic RT simulation, including the effects of both dust extinction and emission, to obtain correct rest-frame $K$-band magnitudes. Lastly, the case study presented in the previous paragraph is a good example of how the data in the extended public database can help narrow down a selection of EAGLE galaxies of interest. This selection can then be studied in more detail by performing RT post-processing with a slightly adjusted configuration as described in \autoref{pythonframework.sec}.

\subsection{Model variations}
\label{ModelVariations.sec}

As indicated in \autoref{Uncertainties.sec}, our post-processing model uses a fixed value for the dust-to-metal fraction, $f_\mathrm{dust}=0.3$, irrespective of galaxy type or redshift. This value is consistent with the observed range from 0.2 to 0.4 \citep{Issa1990, Dwek1998, Watson2011, Brinchmann2013, Zafar2013}. At the same time, recent observations suggest that the dust fraction for galaxies with low gas-phase metallicities varies significantly with metallicity. For example, observations of nearby galaxies \citep{Remy-Ruyer2014,Remy-Ruyer2015} indicate that the dust fraction rapidly increases for metallicities below $0.2~\Zsun$, and continues to gradually increase for higher metallicities. At higher redshifts, there is more uncertainty. Observations of damped Lyman-$\alpha$ absorbers up to $z\approx5$ \citep{Khare2012, DeCia2013, Wiseman2017} suggest that low-metallicity systems at higher redshifts have lower dust fractions than typical nearby galaxies. However, it is not clear whether the dust fraction of the absorbers can directly be assumed to be the same as the dust fraction of galaxies observed in emission. Theoretical models have also suggested an evolution in the dust-to-metal fraction of galaxies as a function of metallicity, especially at gas-phase metallicities below $0.2~\Zsun$ \citep{Zhukovska2014,Feldmann2015,Popping2017}.

It is therefore meaningful to probe the impact of a lower dust fraction on our modeled fluxes, especially for low-metallicity and high-redshift galaxies. We handpicked a set of 15 EAGLE galaxies at redshift 5 with metallicities of $Z_\mathrm{gas}\approx 0.05~\Zsun = 0.0006$. The intrinsic stellar mass of the systems varies from 3 to $7\times 10^8~\Msun$ with a SFR of 1 to 4 solar masses per year. The dust to stellar mass ratios, estimated from our fiducial model with $f_\mathrm{dust}=0.3$, range from 3 to $5\times 10^{-3}$, and estimated dust temperatures range from 23 to 30~K.

After re-processing these galaxies using a value of $f_\mathrm{dust}=0.15$ instead of the fiducial value, the estimated dust masses decrease by about 30 per cent. The dust mass does not fully scale with the value of $f_\mathrm{dust}$ because a significant fraction of the dust is modeled by star-forming regions (see \autoref{PostProcessingGalaxies.sec}), while $f_\mathrm{dust}$ affects just the diffuse dust in the model. The estimated dust temperatures barely change (by less than 0.7~K), which is explained by the low optical depth in these systems. The rest-frame UV and optical fluxes increase by 5 to 25 per cent, depending on the galaxy and on the line of sight, because of the diminished dust extinction. The rest-frame mid-infrared ($8~\micron$) flux increases by 25 to 35 per cent, and the continuum dust emission fluxes in the submm wavelength range increase by 30 percent, aligned with the increase in total dust mass.

Similarly, there is significant uncertainty on the value of the PDR covering factor. The value in our fiducial model, $f_\mathrm{PDR}=0.1$, was calibrated to observations of nearby galaxies. The value may, however, increase for high-redshift and more gas rich galaxies. \C\ showed a shift towards colder dust temperatures with increasing $f_\mathrm{PDR}$, caused by the more dispersed obscuration of the star-forming cores by the dust in the PDRs. The effect on the estimated dust mass is similar to the effect of varying the dust-to-metal fraction, caused by the additional dust emission modeled by the star-forming regions. Varying the PDR covering fraction has only a minor effect on optical colors because the dust mass is added in compact regions and does not contribute much to the overall extinction.

Another noteworthy aspect of our post-processing model is the inclusion of dust self-absorption. As mentioned in \autoref{PostProcessingGalaxies.sec}, our code SKIRT takes into account the energy absorbed from dust emission (``self-absorption'') as well as the energy absorbed from stellar emission. Because the self-absorbed energy in turn affects the dust emission pattern, this is an iterative process. The iteration is considered to converge when the total absorbed dust luminosity is less than one per cent of the total absorbed stellar luminosity, or when the total absorbed dust luminosity has changed by less than three per cent compared to the previous iteration. To evaluate the importance of this computationally demanding iteration, we re-processed a handpicked set of EAGLE galaxies ignoring dust self-absorption (i.e.\ taking into account dust absorption from stellar emission only).

Because the fluxes calculated for galaxies requiring many iterations are likely to be affected by dust self-absorption, we selected all galaxies from the RefL0100N1504 EAGLE model that require 8, 9 or 10 self-absorption iterations (no galaxy in the model requires more than 10 iterations). The resulting set contains 65 galaxies. For these galaxies, ignoring dust self-absorption underestimates the dust mass by up to 15 per cent, and the dust temperature by 3 to 8~K. The rest-frame continuum dust emission fluxes are underestimated significantly as well. The largest discrepancies, up to a factor of 2.5, are shown in the 24 to 100~$\micron$ wavelength range, which is compatible with the estimated dust temperatures. These results underline the importance of including self-absorption in the post-processing procedure.

While most galaxies in our high-self-absorption selection are at high redshifts ($2\lesssim z\leq5$), some are at redshifts down to $z=0.3$. This is surprising, because galaxies at higher redshifts are more likely to be both compact and highly active, which can provide the high optical depths and high dust temperatures that lead to significant dust self-absorption. It appears that some EAGLE galaxies at lower redshift share these properties as well. Indeed, all galaxies in the set are fairly massive ($M_*\gtrsim10^{10}~\Msun$) and active ($\mathrm{SFR}\gtrsim 20~\Msun\,\mathrm{year}^{-1}$), contain a fair amount of dust ($M_\mathrm{dust}/M_*\gtrsim10^{-3}$), and show representative dust temperatures above 30~K (estimated with dust self-absorption enabled). However, these properties do not set the selected galaxies apart from galaxies with less prominent dust self-absorption: the RefL0100N1504 EAGLE model contains over 750 galaxies that satisfy these criteria. A likely conclusion is that the amount of self-absorption heavily depends on the specific geometry of a galaxy, so that it is impossible (or at least nontrivial) to predict whether a particular galaxy requires the self-absorption treatment without actually performing the procedure.


\section{Conclusions}
\label{Conclusions.sec}

The EAGLE project \citep{Schaye2015, Crain2015} consists of a suite of SPH simulations that follow the formation of galaxies and large-scale structure in cosmologically representative volumes. The existing public EAGLE database \citep{McAlpine2016} offers intrinsic properties for galaxies in the EAGLE simulations or ``models", for 29 snapshots at redshifts ranging from $z=20$ to present-day. In this work, we extend the public database with dust-attenuated and dust emission photometry in 50 bands from UV to submm for 316\,389 sufficiently resolved EAGLE galaxies, residing in 23 redshift bins up to $z = 6$, for the six most widely studied EAGLE models. The selection criteria for including an EAGLE galaxy in the extended data set include a minimum stellar mass ($M_*>10^{8.5}~\Msun$) and a minimum number of numerical particles representing the dust content in the galaxy ($N_\mathrm{dust}>250$). This selection excludes some massive galaxies with little dust.

We describe our method to post-process the EAGLE galaxies using the RT transfer code SKIRT \citep{Baes2011, Camps2015a}, essentially following the procedure set forth by \citet{Camps2016} and \citet{Trayford2017}. The procedure handles specific components for star formation regions, stellar sources, and diffuse dust, takes into account stochastic heating of dust grains, and self-consistently calculates dust self-absorption. We assume fixed dust properties, including a fixed dust-to-metals ratio, at all redshifts. We apply the appropriate redshift and filters to the simulated SEDs to obtain broad-band photometry corresponding to astronomical instrumentation. We estimate that the numerical uncertainty on the calculated magnitudes due to our post-processing procedure is $\pm0.05$~mag. \autoref{tables.tab} and \autoref{bands.tab} describe the extra fields in the extended database. We also publish the Python framework implementing our procedures as open source software. Given that the complete data for all EAGLE snapshots are publicly available \citep{EAGLEteam2017}, this allows any third party to reprocess EAGLE galaxies with an adjusted parameter configuration, for example to produce full data cubes or images rather than spatially integrated quantities.

We report a number of checks of the newly published data, from which we conclude that the results generally match expectations. For example, we look at some stacked SEDs (\autoref{StackedSEDs.fig}), we evaluate the accuracy of SFR indicators using NUV and 24~\micron\ fluxes for present-day galaxies (\autoref{StarFormationRate.fig}), we estimate the dust mass and temperature from the submm fluxes in the database, and we plot several dust-related relations at multiple redshifts. These relations include dust mass versus stellar mass (\autoref{DustStellarMass.fig}), optical reddening versus dust mass (\autoref{ColorsMultiRedshift.fig}), and submm color $f_{250}/f_{350}$ versus $f_{350}/f_{500}$ (\autoref{SubmmColorsRedshiftZero.fig} and \autoref{ColorsMultiRedshift.fig}). We also study contributions from dust attenuation and emission in the $K$-band (\autoref{KbandDiscrepancy.fig}). Our results show that dust emission can contribute significantly to the rest-frame $K$-band luminosity, especially at higher redshifts.

Using this newly published set of dust-aware simulated galaxy photometry, it becomes possible to compare yet another aspect of the EAGLE models with observations. For example, we plan such comparisons with observations by the \emph{Herschel} Astrophysical Terahertz Large Area Survey \citep[H-ATLAS;][]{Eales2010} up to redshift $z=0.5$. More specifically, we would attempt to reproduce the evolution of a number of properties of the galaxy population for both optically and submm selected samples \citep[e.g.,][]{Dunne2011, Bourne2012, Bond2012, Smith2012}. We may also further investigate the $K$-band contribution of dust emission and extinction. Other authors have indicated their intent to study the SFR-stellar mass relation for higher redshifts and the various SFR indicators in use, in an attempt to help clarify the tension between observed results, especially at redshifts $z\gtrsim1$ \citep{Bauer2011,Katsianis2016,Katsianis2017}. There are many more possible areas of study, and we invite interested readers to employ the published data in any way they see fit. This research may lead to some insights in the underlying physical processes, and should at least help map the successes and limitations of our numerical models and inform the design of future cosmological simulation projects.


\acknowledgments

\section*{Acknowledgments}

This work fits in the CHARM framework (Contemporary physical challenges in Heliospheric and AstRophysical Models), a phase VII Interuniversity Attraction Pole (IAP) programme organised by BELSPO, the BELgian federal Science Policy Office. This work used the DiRAC Data Centric system at Durham University, operated by the Institute for Computational Cosmology on behalf of the STFC DiRAC HPC Facility (www.dirac.ac.uk). This equipment was funded by BIS National E-infrastructure capital grant ST/K00042X/1, STFC capital grants ST/H008519/1 and ST/K00087X/1, STFC DiRAC Operations grant ST/K003267/1 and Durham University. DiRAC is part of the National E-Infrastructure. This research was supported in part by the Netherlands Organization for Scientific Research (NWO) through VICI grant 639.043.409. RAC is a Royal Society University Research Fellow.

\bibliography{eaglesurv}




\end{document}